\documentclass{aa}  

\usepackage{array,tabularx}
\usepackage{graphicx}
\usepackage{tabularx}
\usepackage{makecell}
\usepackage{txfonts}
\usepackage{xcolor}

\usepackage[colorlinks=true, linkcolor=blue, citecolor=blue, urlcolor=blue]{hyperref}

\begin{document}

    \title{The impact of radial migration on disk galaxy star formation histories:}
    \subtitle{II. Role of bar strength, disk thickness, and merger history}

\author{J.P.~Bernaldez\inst{1,2}
  \and I.~Minchev\inst{2}
  \and B.~Ratcliffe\inst{2}
  \and L.~Marques\inst{2}
  \and K.~Sysoliatina\inst{2}
  \and J.~Walcher\inst{2}
  \and S.~Khoperskov\inst{2}
  \and M.~Martig\inst{3}
  \and R.S.~de Jong\inst{2}
  \and M.~Steinmetz\inst{2}
}

\institute{
  Friedrich-Schiller Universit\"at Jena, Jena, Germany\\
  \email{john.paul.bernaldez@uni-jena.de}
  \and
  Leibniz-Institut f\"ur Astrophysik Potsdam (AIP), An der Sternwarte 16, D-14482 Potsdam, Germany\\
  \email{iminchev@aip.de}
  \and
  Astrophysics Research Institute, Liverpool John Moores University, 146 Brownlow Hill, Liverpool L3 5RF, UK
}

   \date{}

   \abstract{
Reconstructing the star formation history (SFH) of disk galaxies is central to understanding their growth and evolution, yet such estimates can be strongly biased by stellar radial migration over cosmic time. Using 186 Milky Way (MW) and Andromeda (M31) analogs from the TNG50 cosmological simulation, we compare star formation rates (SFRs) inferred from present-day stellar positions with those based on stellar birth radii to quantify the magnitude, spatial structure, and temporal evolution of migration-induced biases. We find that radial migration systematically produces artificial star formation in regions that had not yet formed stars. Notably, $\sim80\%$ of galaxies exhibit outer-disk stars older than 10~Gyr, which we find to have formed at radii interior to the outer disk and to have reached their present locations via substantial outward migration. Similar effects appear in $\sim45\%$ of galaxies at intermediate radii during early epochs, and in 30\% of quenched inner disks within the past 4~Gyr. Migration also smooths SFHs, washing out localized bursts and suppressions by dispersing stars across neighboring radii. The strength and imprint of these distortions depend sensitively on galactic structure and evolutionary history: strong bars drive mean SFR overestimates of up to 75\% in the inner disk and 150\% in the outskirts; thinner, dynamically cold disks suffer average outer-disk biases up to 160\%; while thick disks exhibit typical inner-disk biases up to 125\%. Merger timing further modulates these patterns. Our results demonstrate that failing to account for stellar migration can lead to severe misinterpretations of when and where stars formed, with direct implications for the chemical and evolutionary histories of the MW and external galaxies.
}
   \keywords{}
    \authorrunning{Bernaldez, Minchev, et al.}
    \titlerunning{Radial migration effects on disk galaxy SFHs}
   \maketitle

\section{Introduction}
The primary objective of galactic archaeology is to understand the formation and evolutionary history of galaxies by analyzing the properties of their stellar populations \citep{Eggen1962,Freeman2002,Helmi2020, Rix2013}. One of the most powerful tools in this field is the star formation history (SFH), which provides a timeline of how stars have formed within different regions of a galaxy \citep{Tinsley1980}. Specifically, the SFH is the record of how the star formation rate (SFR), which is the mass of new stars formed per unit time per unit area, has varied across different regions and epochs of a galaxy's evolution \citep{Kennicutt2012}. Recovering a galaxy's SFH is fundamental to understanding the physical mechanisms behind its growth, including gas accretion, stellar feedback, and external influences, such as mergers and interactions \citep{Madau2014, Matteucci1989, Weisz2014}. By studying the SFH, we can identify periods of intense star formation, triggered either by internal dynamical processes or external environmental factors, as well as episodes of quenching when star formation ceased due to gas depletion or feedback effects \citep{Tacchella2016, khoperskov18}. 

Analyses of the SFH of very nearby galaxies have traditionally relied on color-magnitude diagrams (CMDs), which compare observed stellar distributions with synthetic ones, an approach particularly effective for dwarf galaxies, where the relatively simple stellar populations allow for detailed reconstructions of their past star formation activity \citep{Gallart1999, Hunter2022, Pan2023, Savino2023}. Additionally, CMD modeling has also been used to study stellar populations and structural properties of galactic bulges \citep{Deras2023} and to trace the Milky Way (MW)’s bulge SFH with Hubble Space Telescope (HST) observations \citep{Bernard2018}, revealing the complex histories of central regions shaped by gas inflows, mergers, and dynamical processes \citep{Kormendy2004,Noguchi1999}. Stellar ages from large spectroscopic surveys also complement CMD results by refining MW formation histories \citep{Gallart2019, Gallart2024, fernandez25}. Such CMD analyses are not limited to the MW; studies of M31 offer valuable comparisons between the evolution of the MW and its closest large neighbor \citep{Bernard2015}.

In addition to CMD-based techniques, metallicity gradients have been used in conjunction with chemical evolution models to infer SFHs, offering a complementary approach to understanding galaxy evolution \citep{Daflon2004}. To discuss one of many examples, \cite{Snaith2015} and \cite{Kruijssen2019} employed chemical abundance trends to study the MW's history, building on the finding that the solar neighborhood contains fewer metal-poor stars than expected, suggesting complex formation pathways involving both in situ star formation and the accretion of external material \citep{Schmidt1963, Tinsley1974}. 

Star formation history studies typically assume, often implicitly, that stars remain near their birth radii. While this assumption is reasonable for a galaxy’s total SFH, since radial migration does not change the total stellar mass formed at each epoch when integrated over the full disk \citep{Haywood2013,Kubryk2015}, it becomes problematic for spatially resolved SFHs of disk galaxies, including the MW. It is now well-established that stars in galactic disks do not remain near their birth radii as they are often shifting their guiding radii by several kpc over their lifetimes \citep{Wielen1977, Sellwood2002, Minchev2006, Roskar2008, Athanassoula2013, VeraCiro2014, Frankel2020}.\footnote{Strictly speaking, “radial migration” should refer to a permanent change in guiding radius (i.e., angular momentum). Stars with hotter kinematics, however, will also be found away from their birth radii because of epicyclic “blurring” or the asymmetric drift effect \citep{bt08}, which, in principle, can be corrected for when kinematic information is available. Such blurring typically displaces stars only moderately from their birth places, but it can nevertheless bias spatially resolved SFHs. Throughout this work we use the term “radial migration” more loosely to refer to both effects, i.e., to any sustained radial mismatch between present-day and birth radii.}
This angular momentum redistribution is driven by various dynamical processes, including transient spiral arms \citep{Roskar2008, Grand2012}, the overlap of multiple non-axisymmetric patterns \citep{Minchev2010, Minchev2011, vislosky24, Marques2024, kwak26a}, rapid bar slow down after formation \citep{khoperskov20, zhang25}, and external perturbations such as satellite interactions or minor mergers \citep{Quillen2009,bird12}. In addition to large-scale drivers, scattering off gas clouds, clumps, and disk-halo flows can redistribute stellar orbits and help maintain exponential disk profiles \citep{elmegreen16, struck17, struck18, wu20}. These processes may be particularly important in dwarf irregulars, where star formation appears outside-in \citep{zhang12}, and in the outer regions of disks, where spirals and bars are less important.

Observations and simulations suggest that the strength of radial migration can vary significantly, depending on the galaxy history, the strength of the perturbers, the disk radius, and the stage of its evolution. Observationally, studies of the MW's disk using RAVE \citep{kordopatis15}, APOGEE \citep{Marques2024}, and Gaia \citep{nepal24} data have revealed that metal-rich stars in the solar neighborhood likely originated from the inner disk, as predicted by MW chemo-dynamical models \citep{Kubryk2015,Minchev2013.1}, while more metal-poor stars have formed farther out \citep{Frankel2018,Hayden2015}, in agreement with a negative metallicity gradient. The typical strength of radial orbit migration has been estimated to be approximately $3.6\pm0.1$ kpc, based on an action-based parametrized model \citep{Frankel2020}, a result supported by more recent modeling that accounts for the time evolution of the star-forming region and bar strength effects \citep{Ratcliffe2025}. However, \citet{verma21} found, unsurprisingly, that the amplitude of migration depends on age, with values of $\sim 2 \,\mathrm{kpc}$ for young ($<4 \,\mathrm{Gyr}$) and $\sim 3 \,\mathrm{kpc}$ for older ($4$--$8 \,\mathrm{Gyr}$) stellar populations in the solar cylinder. These migration amplitudes are also consistent with estimates derived from the Auriga simulations \citep{Okalidis2022}.

Many works have explored the chemo-dynamical evolution of the MW disk, demonstrating how radial migration influences the chemical and kinematic properties of stars across different Galactic regions (e.g., \citealt{johnson25,Kubryk2015, Minchev2014, prantzos23,Schonrich2009,sharma21}). In particular, these studies highlight how migration affects metallicity gradients and the distribution of stellar populations over time. \citet{Ratcliffe2023} further refined our understanding of this process by unveiling the time evolution of chemical abundances across the MW disk using APOGEE data, offering crucial observational constraints on radial migration models. This chemical evidence aligns with the predictions of radial migration models, reinforcing the need to account for this process when reconstructing the MW's SFH \citep{Francois1993,Ratcliffe2024b}. While observational studies can infer birth radii and stellar migration using chemo-kinematic modeling \citep{Minchev2018, Ratcliffe2023, Ratcliffe2024}, a key advantage of simulations lies in their ability to directly track the birth locations and dynamical evolution of stars in a self-consistent framework across cosmic time. This is particularly important because SFHs reconstructed from present-day stellar distributions implicitly assume that stars formed near their current locations, leading to biased results.
In the first paper of this series, \cite{minchev25} quantified how radial migration biases spatially resolved SFHs using eight disk galaxies simulated in the cosmological context, as described by \cite{Martig2012}. They found that reconstructions based on present-day stellar positions underestimate early star formation in the inner disk by 25--50\% and overestimate recent activity, while in the very central regions ($\sim0.4h_d$) SFRs are consistently inflated due to inward migrators. At intermediate radii and in the outskirts, genuinely in situ populations are underestimated by $\sim50\%$, whereas migrated stars from the inner disk boost apparent SFRs by up to 200\%. The net effect is that SFR peaks are suppressed and broadened, and the true rate of inside-out growth is systematically underestimated. Because migration depends on both radius and time, they noted that correcting for these biases in external galaxies is challenging without additional constraints such as disk mass, bar strength, or vertical-to-radial scale ratios.

In this second paper of the series, we extend the analysis of \citet{minchev25} using the TNG50 cosmological simulation \citep{Nelson2019CAC, Pillepich2024}, which provides a much larger sample of MW- and M31-mass galaxies. Whereas the small sample in \citet{minchev25} did not allow a statistical assessment of how SFH reconstruction biases depend on galaxy morphology or assembly history, the TNG50 sample enables us to quantify, in a systematic way, how migration-induced distortions vary with bar strength, disk thickness, and the mass and timing of the most significant merger. We do so by tracing both birth and present-day radii of stellar particles and comparing in situ and reconstructed SFHs across different radial ranges.

This paper is structured as follows. Section~\ref{sec:TNG50} discusses the TNG50 simulations and galaxy selection. Section~\ref{sec:Classification} details the classification scheme and compares SFH distortions across galaxy types. Section~\ref{sec:methods} describes the methods used to reconstruct the SFH and Section~\ref{sec:results} discusses the results of the analysis. Section~\ref{sec:Conclusion} summarizes the key findings and their implications.

\section{TNG50 MW and M31 analogs}
\label{sec:TNG50}

The IllustrisTNG project \citep{Marinacci2018,Naiman2018,Nelson2018,Nelson2019CAC, Pillepich2018,Springel2010, Springel2018} is a set of high-resolution large cosmological magneto-hydrodynamical simulations of galaxy formation in the context of the $\Lambda$CDM (Lambda Cold Dark Matter) cosmological model \citep{Planck2016}. The TNG50 simulation \citep{Nelson2019,Pillepich2019} has the highest resolution in the TNG project and is the most computationally demanding since it is designed to model the processes involved in the formation of galaxies, stars, and black holes. The TNG50 simulation traces the evolution of various cosmic components, including cold dark matter, gas, stars, supermassive black holes, and magnetic fields, within a cubic volume measuring 51.7 comoving Mpc per side sampled by $2160^3$ gas and dark matter cells. This simulation spans from a redshift of $\mathit{z}=127$ which corresponds to a time shortly after the Big Bang, up to the present day at $\mathit{z}=0$. 

TNG50 includes a vast range of galaxies over different environments, from massive galaxies residing at the centers of or orbiting within galaxy clusters with masses as high as $10^{14.3}M_\odot$, to isolated dwarf galaxies or those found in smaller groups. Among these thousands of galaxies are 198 MW- and M31-like galaxies at redshift $\mathit{z}=0$ \citep{Pillepich2024}. These galaxies have stellar masses within the range $\text{M}_*(<30 \text{kpc})=10^{10.5-11.2} \text{M}_\odot$, exhibit disk-like morphologies, including the presence of spiral arms, which is identified by either visual inspection of the galaxies or from the minor-to-major axis ratio of the galaxy's stellar mass distribution. Lastly, they are isolated, with no other galaxies having a stellar mass $\ge 10^{10.5} \text{M}_\odot$ within 500 kpc of their centers, and the total mass of their host halos is less than $\text{M}_{200c}(\text{host})<10^{13} \text{M}_\odot$, where $\text{M}_{200c}$ denotes the mass enclosed within a radius where the average density is 200 times the critical density of the Universe at that time. 

In what follows we adopt the standard TNG50 nomenclature and refer to these systems as “MW-like” and “M31-like” galaxies, as defined in the TNG50 catalogs \citep[e.g.,][]{Pillepich2024}. This label, however, reflects a broad selection in stellar mass, morphology, environment, and halo mass, rather than a detailed match to the specific properties or merger histories of the real MW or M31; indeed, current constraints indicate that the recent merger histories of the MW and M31 are quite different from one another \citep[e.g.,][]{hammer18}. As a result, the sample spans a wide range of bar strengths, gas fractions, and merger histories, including a broad distribution of last major merger times. This diversity is desirable for our purposes, since we use the TNG50 MW- and M31-like population statistically to study how migration-induced SFH biases depend on structural and assembly properties within the MW- and M31-mass regime, rather than to model the individual histories of the MW or M31.

To study each galaxy's evolution, we used 89 simulation snapshots, which record the positions and properties of stellar particles at different cosmic times. In TNG50, snapshots are spaced unevenly but typically cover intervals of around 100–200 million years. Among the 198 galaxies, three were excluded because they contained no stars before 2.8 Gyr, preventing early star formation and migration analysis. Additionally, nine galaxies were removed due to unstable disk alignment across simulation snapshots, identified by irregular changes in the disk orientation between consecutive snapshots, particularly in edge-on projections. Such misalignments, typically resulting from past major mergers, lead to warped or disrupted stellar disks and prevent consistent rotational alignment. From this point on, we analyzed the remaining 186 well-aligned galaxies from the TNG50 simulation, which we used to investigate the effects of radial migration on the SFH for different galaxy classifications.

\section{Classification of galaxies}
\label{sec:Classification}

To understand how different structural and dynamical features influence radial migration and the SFH, we classified the galaxies in our sample according to three key properties: bar strength, disk thickness, and merger history. This allowed us to isolate the effects of each factor and examine how they contribute to the diversity of evolutionary paths in MW– and M31–like systems. In the following subsections, we describe the criteria and methods used to group galaxies by these properties.

 \subsection{Bar strength}
\label{subsubsec:BarStrength}

\begin{figure}
   \centering
   \leftskip-.1cm 
   \includegraphics[width=\columnwidth]{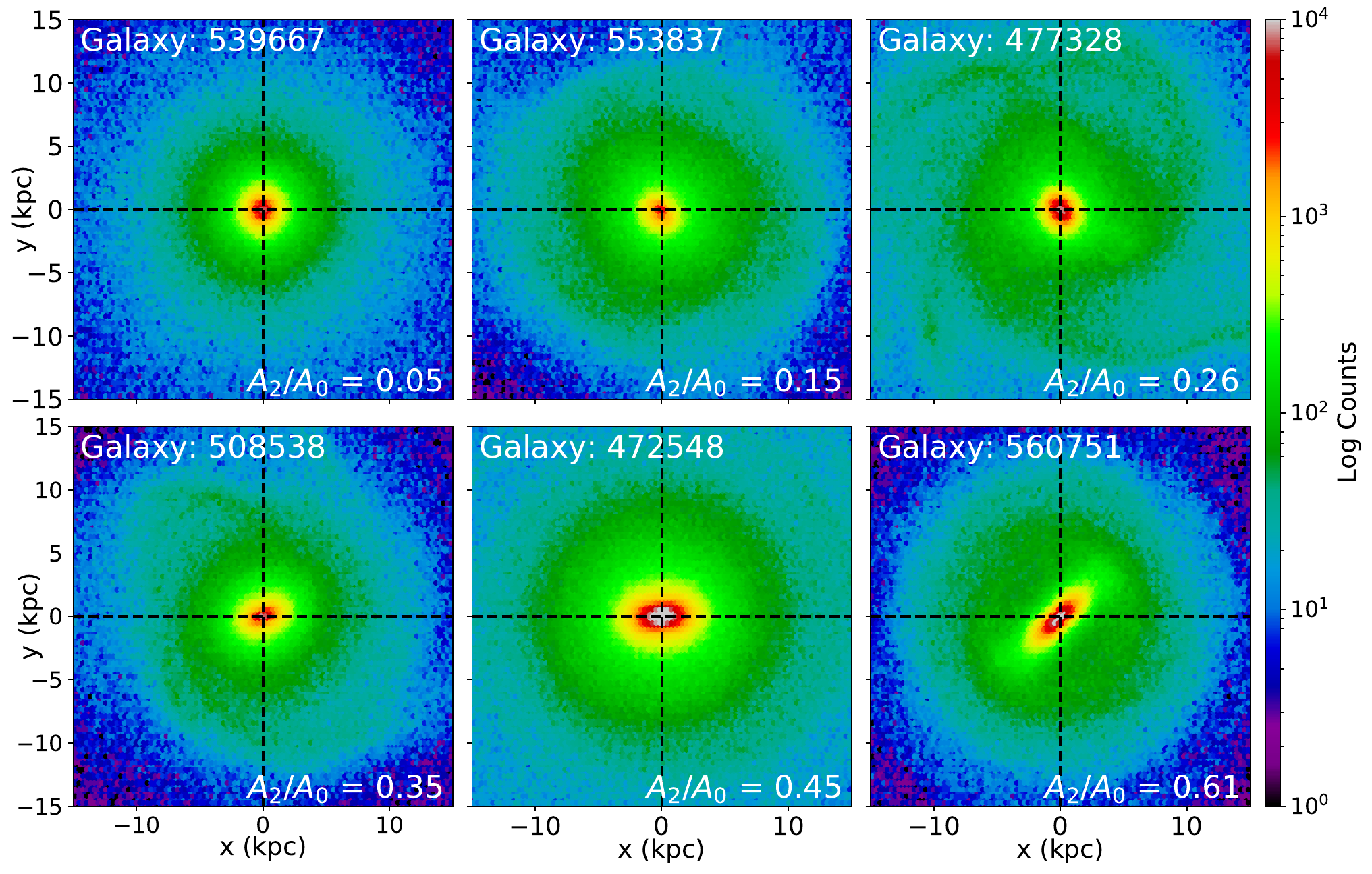}
   \caption{Comparison of galaxy stellar density maps across different bar strength bins, using bar strength measurements from \cite{Khoperskov2024}. In each panel, galaxies are shown face-on. Bar strength increases from left to right and from top to bottom. Galaxies with weak bars (top panels) appear more symmetric and round, while galaxies with stronger bars (bottom panels) clearly show more elongated and asymmetric central structures, characteristic of prominent bars.}
   \label{Fig_BarredGalaxy}
\end{figure} 

Bars are elongated stellar structures that extend from the central regions of disk galaxies and play a fundamental role in secular evolution by redistributing angular momentum and driving gas inflows toward the central regions \citep{Athanassoula1992,Combes1981,Sellwood2014,Sheth2005}. They are observed in approximately 60-70\% of massive disk galaxies in the local Universe \citep{Eskridge2000, Masters2011}. By funneling gas inward, bars can trigger central starbursts, contributing to the growth of bulges, but they can also suppress star formation by rapidly depleting cold gas reservoirs or driving outflows \citep{Cheung2013, Gavazzi2015, semczuk24}. The extent of this influence depends on various bar parameters, such as bar strength, length, and pattern speed, which govern how efficiently bars redistribute stars and gas through radial migration, thereby shaping both the chemical evolution and the SFH of galaxies \citep{Geron2023, Marques2024,Minchev2013.1,Minchev2014}. Stronger bars are particularly associated with enhanced radial mixing, which alters the chemical and dynamical properties of a galaxy's disk by redistributing stars and gas over large distances \citep{Fragkoudi2020}. However, numerical simulations, including those from the IllustrisTNG project, indicate that the efficiency of bar-driven processes, such as resonant scattering, bar-induced gas inflow toward the central regions, and the associated redistribution of stellar orbits, depend not only on bar strength but also on factors such as gas content, dark-matter halo properties, and disk stability \citep{Peschken2019, Rosas-Guevara2022}. These findings suggest that bars are complex dynamical structures that can both promote and regulate star formation, making them a key mechanism in the long-term evolution of disk galaxies.

We adopted the bar strength measurements for each galaxy in the TNG50 simulation from \cite{Khoperskov2024}. These bar strengths, denoted as $A_2/A_0$, represent the peak value of the $m=$ 2 Fourier harmonic of the stellar surface density, providing a quantitative measure of the prominence of bar-like structures within the galaxies. The value is measured in the inner galaxy, within radii where the position angle of the stellar surface-density major axis changes by less than $5^\circ$ over $\Delta R = 0.5\,\mathrm{kpc}$ \citep{athanassoula02}.

A higher $A_2/A_0$ value indicates a stronger and more well-defined bar, whereas lower values correspond to galaxies with weak or no bars. The galaxies from the TNG50 simulation were categorized into six bins based on their bar strength, ranging from 0.0 to 0.6 in increments of 0.1, with the final bin including all galaxies with bar strength values greater than 0.5. This classification allowed us to systematically analyze the effects of varying bar strengths on galactic morphology. 

Figure \ref{Fig_BarredGalaxy} shows a representative galaxy from each bar strength bin, illustrating how their structural properties change with increasing $A_2/A_0$. The top-left panel features galaxy ID 539667, which has the weakest bar strength at $A_2/A_0=0.05$. When viewed face-on, this galaxy appears nearly axisymmetric, with a smooth and circular central region, indicating an absence of any prominent bar. Similarly, the top-center panel shows galaxy ID 553837, with a bar strength of $A_2/A_0=0.15$. Despite slight irregularities in the central mass distribution, no clear bar structure is evident in this case either. These two examples demonstrate that galaxies with very low bar strengths tend to exhibit nearly symmetrical disk structures without elongated central features. The top-right and bottom-left panels show galaxies with intermediate bar strengths: galaxy IDs 477328 and 508538, which have medium bar strengths, of $A_2/A_0=0.26$ and 0.35 respectively. In contrast to the galaxies in the lowest bin, these galaxies exhibit small but noticeable elongated structures at their centers. The central asymmetry becomes more pronounced compared to the nearly circular morphology of the previous examples. The emergence of these elongated features marks the transition from a non-barred or weakly barred system to a more structured barred morphology. Finally, the bottom-center and bottom-right panels showcase galaxies with stronger bar strengths: galaxy IDs 427548 with $A_2/A_0=0.45$ and 560751 with $A_2/A_0=0.61$. Among these, galaxy ID 560751 possesses the strongest bar, displaying a well-defined, elongated structure extending from the galactic center into the outer disk. The bar is significantly more prominent than in the previous cases, spanning a large fraction of the galaxy's central region.

\subsection{Disk thickness}
\label{subsubsec:DiskThickness}

We further categorized the galaxies based on their disk thickness, a crucial structural property that influences galactic dynamics and evolution. This classification was based on the ratio of the thick disk scale-height ($h_{z,thick}$) to the disk scale-length ($h_d$), as calculated in Appendix \ref{app:ScaleLengths}. Figure \ref{Fig_DiskRatio} shows the ratios $h_{z,thick}/h_d$ and $h_{z,thin}/h_d$. The overlaid histogram shows the distribution of these values, while red vertical dashed lines at 0.3 and 0.46 mark the boundaries used to classify disks into thinner, intermediate, and thicker categories based on their vertical-to-radial scale ratios. The tighter clustering of thin disk values (on the y-axis) compared to the broader spread in thick disk ratios suggests more uniformity in thin disk structures. Furthermore, the overall positive correlation between $h_{z,thick}/h_d$ and $h_{z,thin}/h_d$ indicates that galaxies with thicker thick disks also tend to host vertically more extended thin disk components, suggesting that vertical heating processes affect both disk components to some extent. By examining their ratio, we could systematically distinguish between galaxies with varying degrees of disk thickness.

\begin{figure}
   \centering
   \includegraphics[width= \columnwidth]{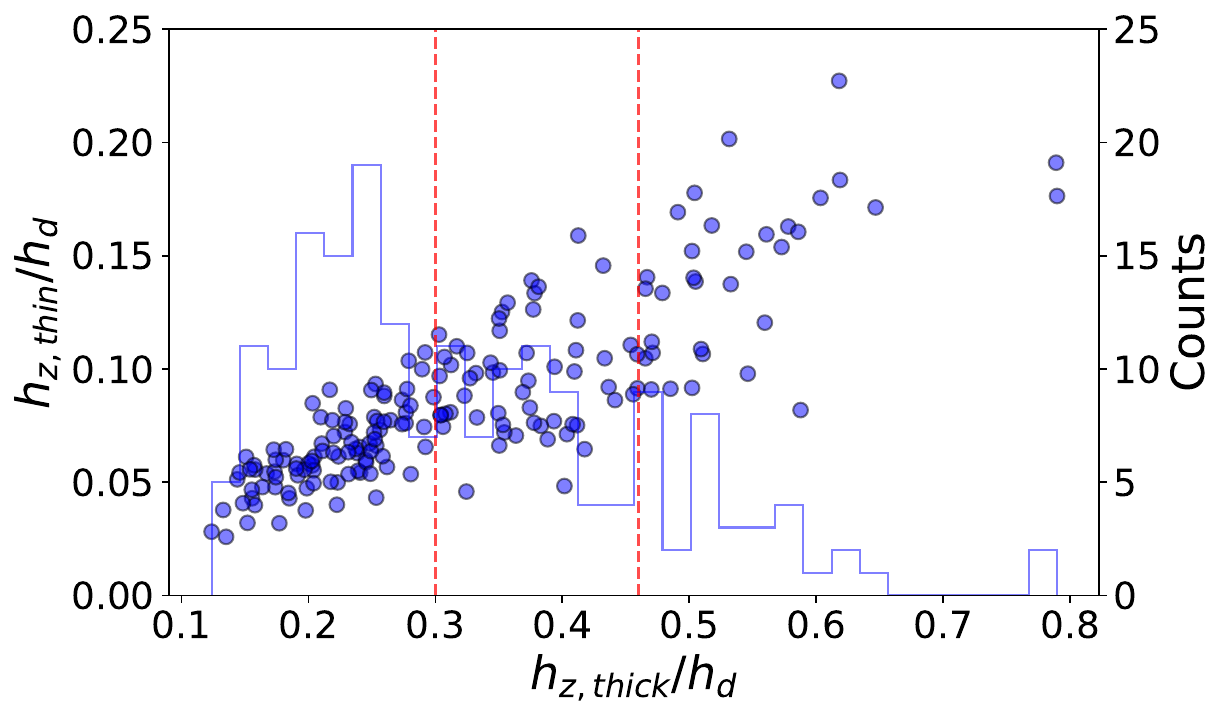} 
   \caption{Comparison between the ratios $h_{z,\text{thick}}/h_d$ and $h_{z,\text{thin}}/h_d$. Disk thickness is classified using $h_{z,\text{thick}}/h_d$: thinner ($\leq 0.3$), intermediate (0.3–0.46), and thicker ($\geq 0.46$) disks, with red dashed vertical lines marking the boundaries. The positive correlation and broader spread in thick disk ratios suggest that vertical heating affects both thin and thick disk components, allowing a systematic distinction between disk thickness categories.}
   \label{Fig_DiskRatio} 
\end{figure}

We defined thicker disk galaxies as those with a scale-height-to-length ratio of $h_{z,thick}/h_d \ge 0.46$. Thicker disk galaxies have a more vertically extended disk structure, which may indicate a history of dynamical heating, interactions, or internal processes that have thickened the disk over time and are expected to host older stars that formed at smaller radii before migrating outward. Such galaxies are often associated with older stellar populations and may exhibit lower rotational support compared to thinner disk systems. Intermediate disk galaxies were defined as those with a ratio in the range $0.30 < h_{z,thick}/h_d < 0.46$. These galaxies represent a transitional class, where the disk is neither extremely thin nor significantly thickened. They may have undergone moderate dynamical heating or could be in the process of evolving toward a thicker or thinner disk state, depending on environmental interactions and internal secular processes \citep{Minchev2014}.

Finally, thinner disk galaxies are those with $h_{z,thick}/h_d \le 0.3$. These galaxies typically exhibit a disk with low vertical velocity dispersions, leading to their characteristically smaller scale-heights compared to their scale-lengths. These galaxies exhibit more flattened disk structures and are associated with younger, metal-rich stellar populations, ongoing star formation, and coherent spiral structure \citep{Yoachim2006}. The preservation of thinner disks over time is often linked to relatively quiescent merger histories, as significant dynamical heating from major mergers tends to thicken disks or disrupt them entirely \citep{Brook2004}. Moreover, stars in thinner disks generally reside on dynamically colder orbits, making them more susceptible to resonant perturbations from non-axisymmetric structures such as bars and spiral arms. As a result, radial migration may proceed more efficiently in these systems, potentially leading to distinct patterns in the reconstructed SFH—particularly in the observed differences between birth and final radius–based estimates of the SFR.

\begin{figure}
   \centering
   \includegraphics[width=\columnwidth]{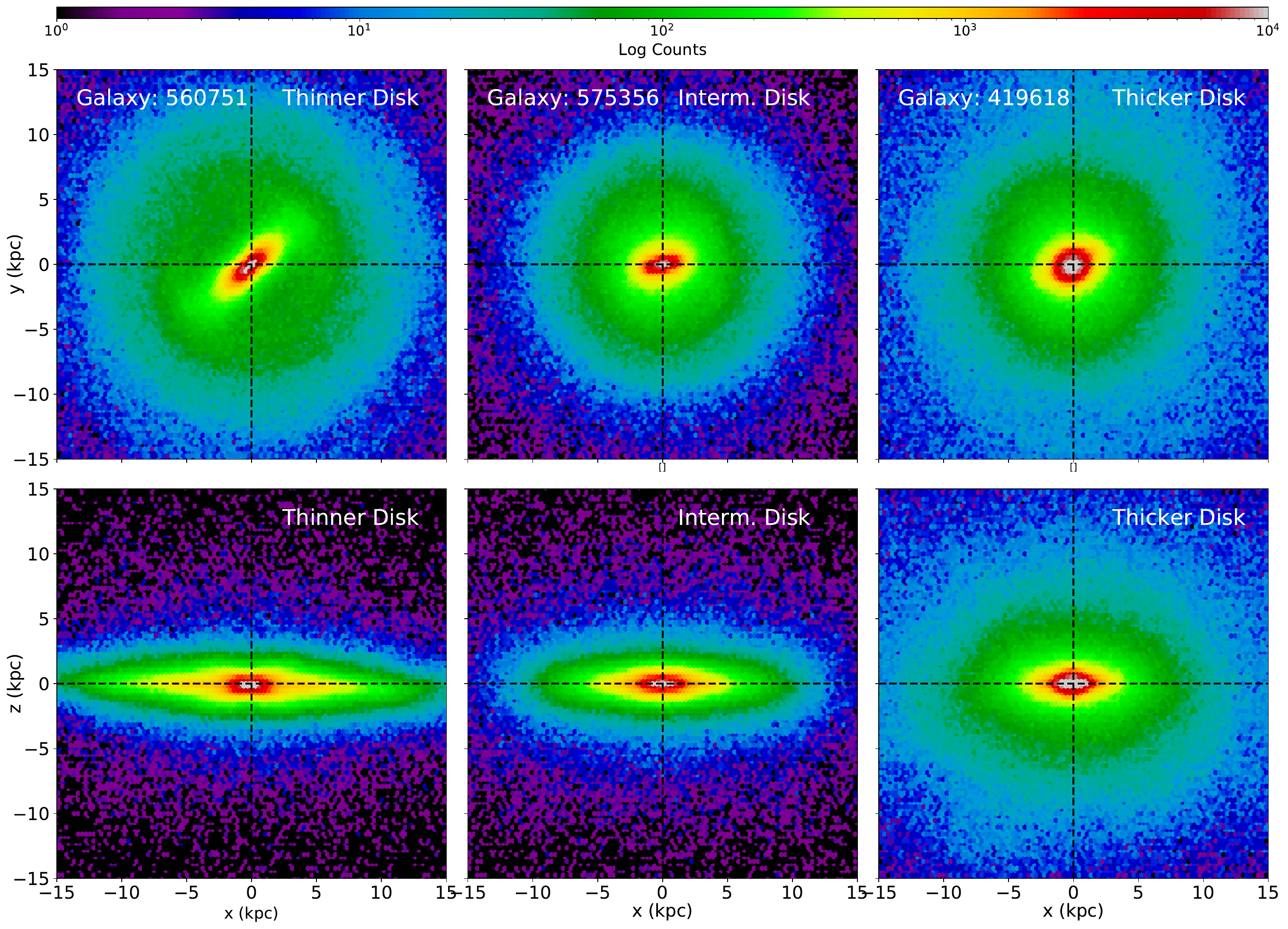}
   \caption{Face-on (top row) and edge-on (bottom row) stellar density maps for three galaxies representing the three disk thickness categories used in this study: a thinner disk (ID: 560751, left), an intermediate thickness disk (ID: 575356, center), and a thicker disk (ID: 419618, right). The vertical extent of each disk becomes progressively larger from left to right in the edge-on views, illustrating the morphological differences captured by the scale-height-to-length ratio $h_{z,thick}/h_d$.}
   \label{Fig:DiskThickness}
\end{figure}

In addition to using the thick disk scale-height for this classification, we also investigated whether the thin disk scale-height, $h_{z,thin}$, could serve as an alternative or complementary structural indicator. The thin disk component generally traces the most recently formed stars and reflects the coldest, most rotationally supported part of the disk. We tested a classification scheme using thresholds based on $h_{z,thin}/h_d$  but found that the resulting groupings were highly correlated with those obtained from the thick disk ratio. In most cases, galaxies identified as thin or thick using the thick disk criterion fell into the same categories when using the thin disk scale-height. With this, we adopted $h_{z,thick}/h_d$  as our primary classification metric.

Figure \ref{Fig:DiskThickness} shows the face-on (top panels) and edge-on (bottom panels) stellar density maps of three galaxies with varying disk thickness. The left panels display galaxy ID 560751, classified as a thinner disk; the middle panels show galaxy ID 575356, representing an intermediate disk; and the right panels present galaxy ID 419618, categorized as a thicker disk. This figure visually illustrates the morphological differences across the disk thickness categories. While all three galaxies appear relatively similar when viewed face-on, the edge-on views reveal significant differences in vertical structure. In particular, the thicker disk (right) exhibits a visibly more extended vertical profile relative to its disk scale length compared to the thinner disk (left), highlighting how disk thickness becomes evident primarily through vertical morphology. This visual confirmation supports the classification based on scale-height-to-length ratios.

\label{subsubsec:Mergers}
\begin{figure}
   \centering
    \includegraphics[width=\columnwidth]{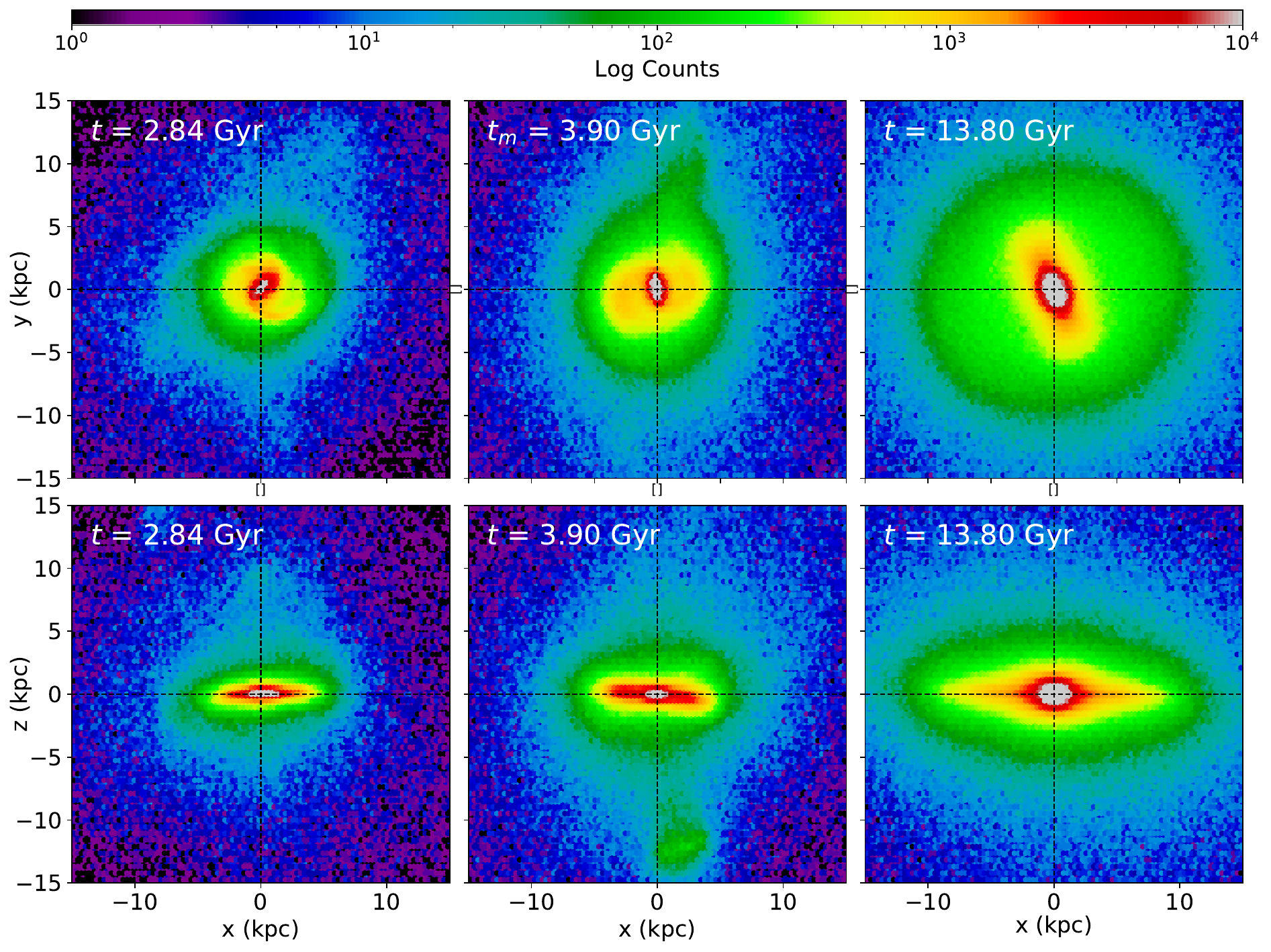}
    \vspace{-.5em} 
   \caption{Example of early merger (galaxy ID 447914), in face-on (upper panels) and edge-on (lower panels) views at three stages in cosmic time: at pre-merger, $t_p=2.84$ Gyr (left), at the time of merger, $t_m=3.90$ Gyr (center), and at final time, $t_f=13.80$ Gyr (right). Initially, the galaxy appears thin and relatively low in mass. Following the merger, the disk becomes more massive and significantly thicker, with stars redistributed to larger vertical distances, suggesting that the merger dynamically heated the disk and contributing to the structural growth of the galaxy.}
    \label{Fig: EarlyMerger}
\end{figure}

\subsection{Merger history}

In the third category, we group galaxies by their merger history. Galaxies in a cosmological context often grow not just through smooth accretion, which is the gradual inflow of diffuse gas from the cosmic web, but also through mergers with smaller satellites. Some of these interactions are relatively minor, while others can leave lasting imprints on morphology, kinematics, and star formation activity \citep{Hopkins2006}. Idealized GSE-like merger experiments from the GASTRO library likewise show that an early clumpy star-forming phase or a retrograde GSE-like merger can significantly reduce the disk SFR, whereas a prograde radial merger has little impact on the global SFR \citep{amarante25}. To better understand how such mergers affect galaxy structure in TNG50, we obtained the SubLink merger trees from the simulation \citep{Nelson2019,Pillepich2018} and identified the most massive merger experienced by each galaxy in our sample.

The most massive merger refers to the event with the highest stellar mass ratio between the infalling progenitor and the host galaxy. For each galaxy, we examined all subhalos that merged with the main progenitor and computed their stellar mass ratios with respect to the host. The snapshot corresponding to the merger with the highest ratio was then selected, and the merger time was recorded by converting that snapshot into look-back time using the cosmology of the simulation. This look-back time served as a reference point to align the SFHs of the galaxies and to classify their merger histories.

Based on this merger time, we grouped galaxies into early, middle, and late merger categories. Galaxies that experienced their most massive merger more than 8 Gyr ago (before 6 Gyr in cosmic time) were classified as early mergers. These events typically occurred while the galaxies were still assembling their disks, and often contributed significantly to the formation of thick disks and central bulges. For instance, galaxy ID 447914 (Figure \ref{Fig: EarlyMerger}) underwent such an early merger. When viewed edge-on, the galaxy appears visibly smaller and thinner prior to the merger. By the final snapshot at $t_f=13.80$ Gyr, the disk has grown substantially, becoming more massive and vertically extended—a structure consistent with early dynamical heating. 

\begin{figure}
   \centering
    \includegraphics[width=\columnwidth]{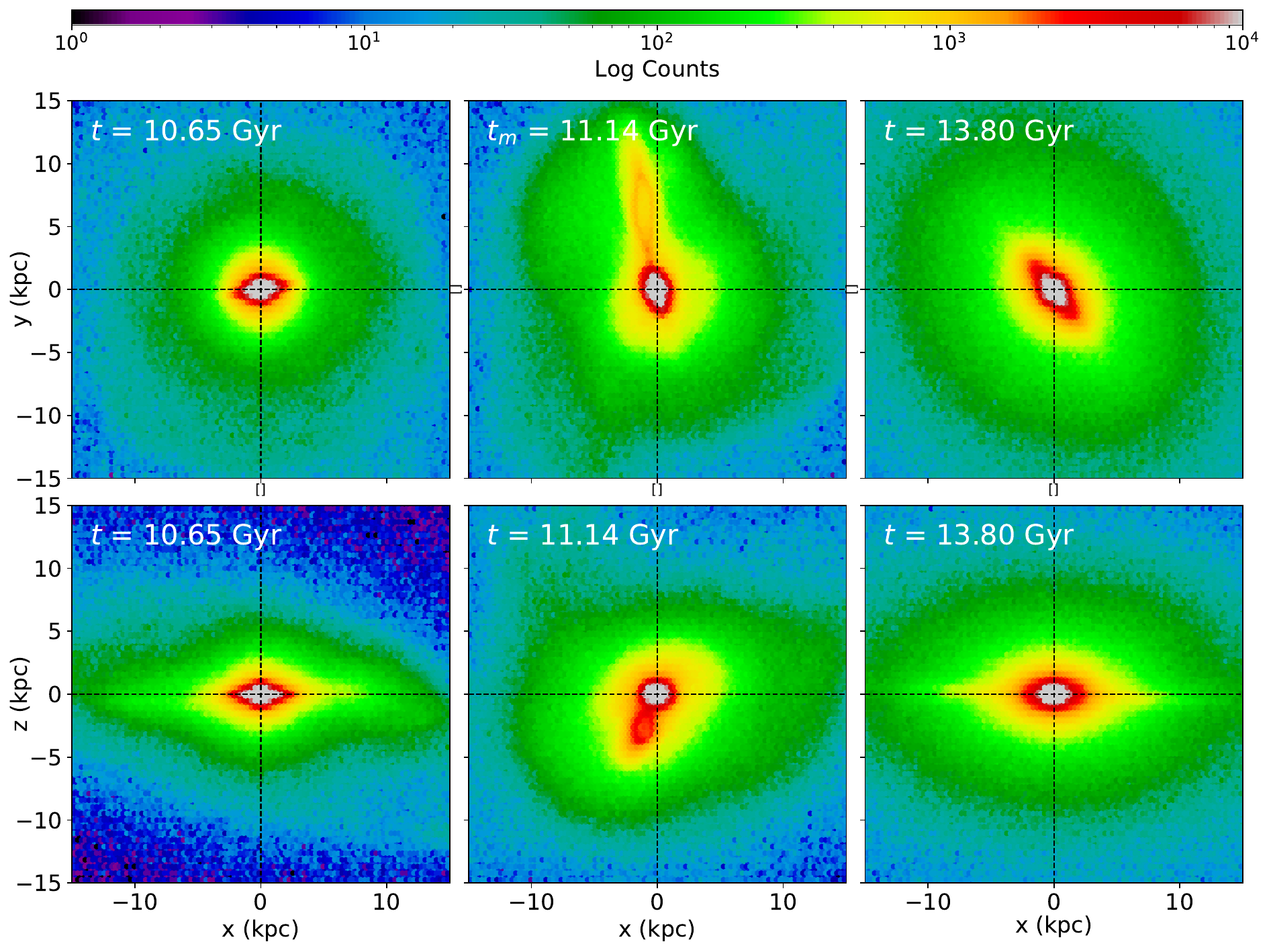}
    \vspace{-.5em} 
    \caption{Example of late merger galaxy (ID: 372754), shown in face-on (top) and edge-on (bottom) views before merger, $t_p=10.65$ Gyr (left), at the time of the merger, $t_m = 11.14$ Gyr (center), and at the final time, $t_f = 13.80$ Gyr (right). At the time of the merger, the galaxy retains some semblance of a disk-like structure, though it already shows signs of disruption. By the final snapshot, the disk becomes more disturbed and puffed up, especially in the vertical direction. This suggests that the late merger deposited energy into the system too recently for the galaxy to fully relax, leading to a disrupted morphology and significant vertical thickening.} 
    \label{Fig: LateMerger}
\end{figure}

Galaxies that experienced their most massive merger within the most recent 4 Gyr (more than 10 Gyr in cosmic time) were classified as late mergers. These recent events tend to be more disruptive, particularly if the merging satellite is relatively massive. Late massive mergers can strongly perturb or even destroy the stellar disk, depending on the orbit and mass ratio of the interaction \citep{Hopkins2009,Kannan2015}. One clear example is galaxy ID 372754 (Figure \ref{Fig: LateMerger}), which underwent a late merger at $t_m=11.14$ Gyr. This event caused a sharp increase in stellar mass and led to the loss of the well-defined disk structure. We also identified a third group of galaxies that have massive mergers occurring between 4 and 8 Gyr ago, or between $6-10$ Gyr in cosmic time. These events typically fall in a transitional regime, where they are not as destructive as late mergers, but can still contribute to disk heating, bulge growth, and shifts in star formation activity. By relating each galaxy's evolutionary history to the time of its most massive merger, we are able to systematically investigate how the timing of massive merger events influence the growth and structure of disk galaxies.

\section{Methods}
\label{sec:methods}
\subsection{Reconstructing the disk SFH}\label{subsec:ReconstructingSFH}

Knowing where a star was born within the galactic disk is key to understanding a galaxy's formation and evolution. By tracing the birth locations of stars, we can examine how they have migrated over time and how this movement shaped the galaxy's present-day structure. The birth radii of all stars in each galaxy were computed from the comoving birth positions stored for each star particle in the simulation. For each star, we used its stellar formation time to identify the two consecutive snapshots that bracket its birth and linearly interpolated the galaxy center between these snapshots to that formation time. We then subtracted this interpolated center from the star’s comoving birth position, converted the resulting galaxy-centered vector to physical units using the corresponding scale factor, and expressed it in galactocentric cylindrical coordinates.

\begin{figure}
   \centering
   \includegraphics[width=\columnwidth]{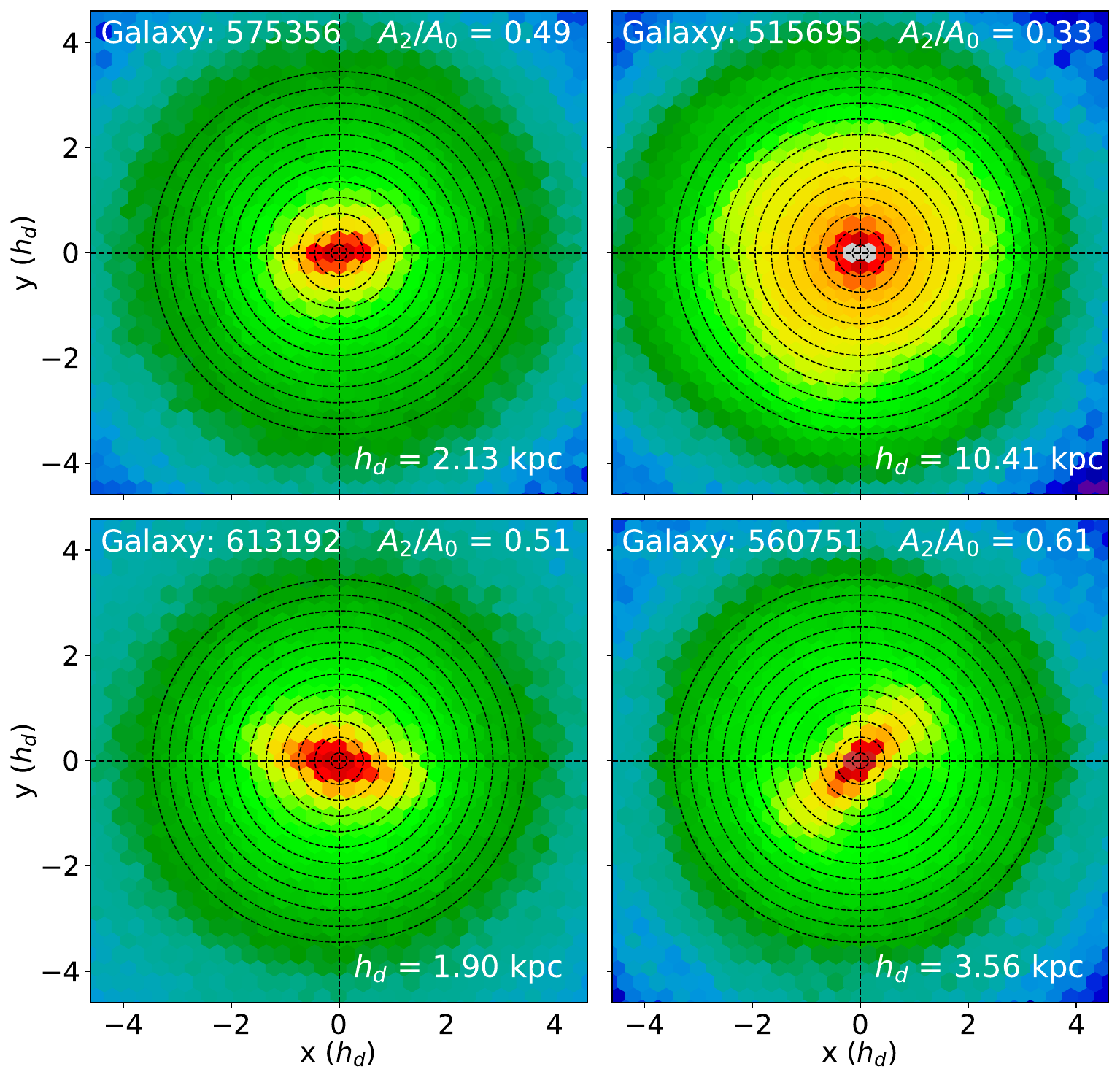}
   \caption{Face-on views of four galaxies at the final time of the simulation, scaled in disk scale-lengths, $h_d$. The concentric circles indicate the 12 radial bins from 0 to 3.6 $h_d$ to show the effects of radial migration.}
    \label{radial_migration}
\end{figure}

In contrast, the final radii, $R_{\text{Final}}$, were determined from the stars' positions, $x_{\text{Final}}$ and $y_{\text{Final}}$, at the last snapshot of the simulation. The birth and final radii were then used to identify the stars for reconstructing the SFH. Specifically, any stars whose birth or final positions lie beyond 4.5 $h_d$, where $h_d$ refers to the disk scale length as defined in Appendix \ref{app:ScaleLengths}, were excluded from the analysis. To quantify the extent of stellar migration over cosmic time, we established 12 equally spaced radial bins, covering a range of $0 < R < 3.6$ $h_d$. Additionally, a vertical position cut was applied, ensuring that only stars within $|z_{\text{Birth}}/h_z|<1$ and $|z_\text{Final}/h_z|<1$ were included, where $h_z$ here refers to the disk scale height thick as obtained from Appendix \ref{app:ScaleLengths}. These restrictions helped focus on stars within the primary disk region, reducing contamination from the halo. Each bin has a width of 0.15 $h_d$, and the distance between the bin centers was 0.3 $h_d$, though results were consistent with different bin sizes. These annuli allowed for a systematic examination of how stars redistribute within the disk. Figure \ref{radial_migration} illustrates four example galaxies, each overlaid with concentric circles corresponding to the radial bin centers used in the SFH reconstruction.

Following \cite{minchev25}, we analyzed the effects of migration on the SFH by calculating the SFR for each radial bin considering both $R_{\text{Birth}}$ and $R_{\text{Final}}$ of stars. The area of each radial bin, $\Delta S$, was determined using the following equation:
\begin{equation}
\centering
    \Delta S = S_n - S_{n-1},
\end{equation}
where $S_n$ represents the area enclosed by the nth concentric circle. For each radial bin, we summed the masses of all stars contained within it to obtain the total stellar mass and extracted their ages. We reconstructed the SFH by computing the SFR in discrete age bins, with a fixed bin width of $1$ Gyr, given by the following equation:
\begin{equation}
    \centering
    \text{SFR} = \frac{M}{\Delta S \times \Delta \text{age}},
\end{equation}
where $M$ is the current stellar mass of all star particles in the given age and radius bin, and $\Delta\mathrm{age}$ is the width of the age bin. Because TNG50 evolves stellar particle masses according to a stellar-evolution mass-loss prescription, this definition yields SFHs that are proportional to the present-day stellar mass formed in each age bin rather than to the instantaneous SFR at the time of formation.
 
To analyze how the stars move from where they are born to present time, we reconstructed the SFH in two ways:
\begin{itemize}
    \item[$\bullet$] SFR$_{\text{Birth}}$ \textendash \ the SFR when binning by $R_\text{{Birth}}$, representing where stars originally formed,
    \item[$\bullet$] SFR$_{\text{Final}}$ \textendash \ the SFR when binning by $R_{\text{Final}}$, representing where stars are observed at the present time.
\end{itemize}
In both cases, we applied the same mass-weighted definition from Eq.~(2), so any age-dependent mass loss affects $\mathrm{SFR}_\mathrm{Birth}$ and $\mathrm{SFR}_\mathrm{Final}$ in the same way; the migration-induced biases we quantify therefore depend only on how the same stars are redistributed between birth and present-day radii.

A smoothed representation of the SFH was achieved by utilizing a rolling approach, where the age bins were shifted by 0.2 Gyr increments, which helped capture trends in star formation while mitigating noise due to stochasticity in the age distribution of stars. 

\subsection{Quantifying the effect of radial migration on the SFH estimates} 
\label{SFH_comparison}

\begin{figure*}
   \centering
   \includegraphics[width=2.0\columnwidth]{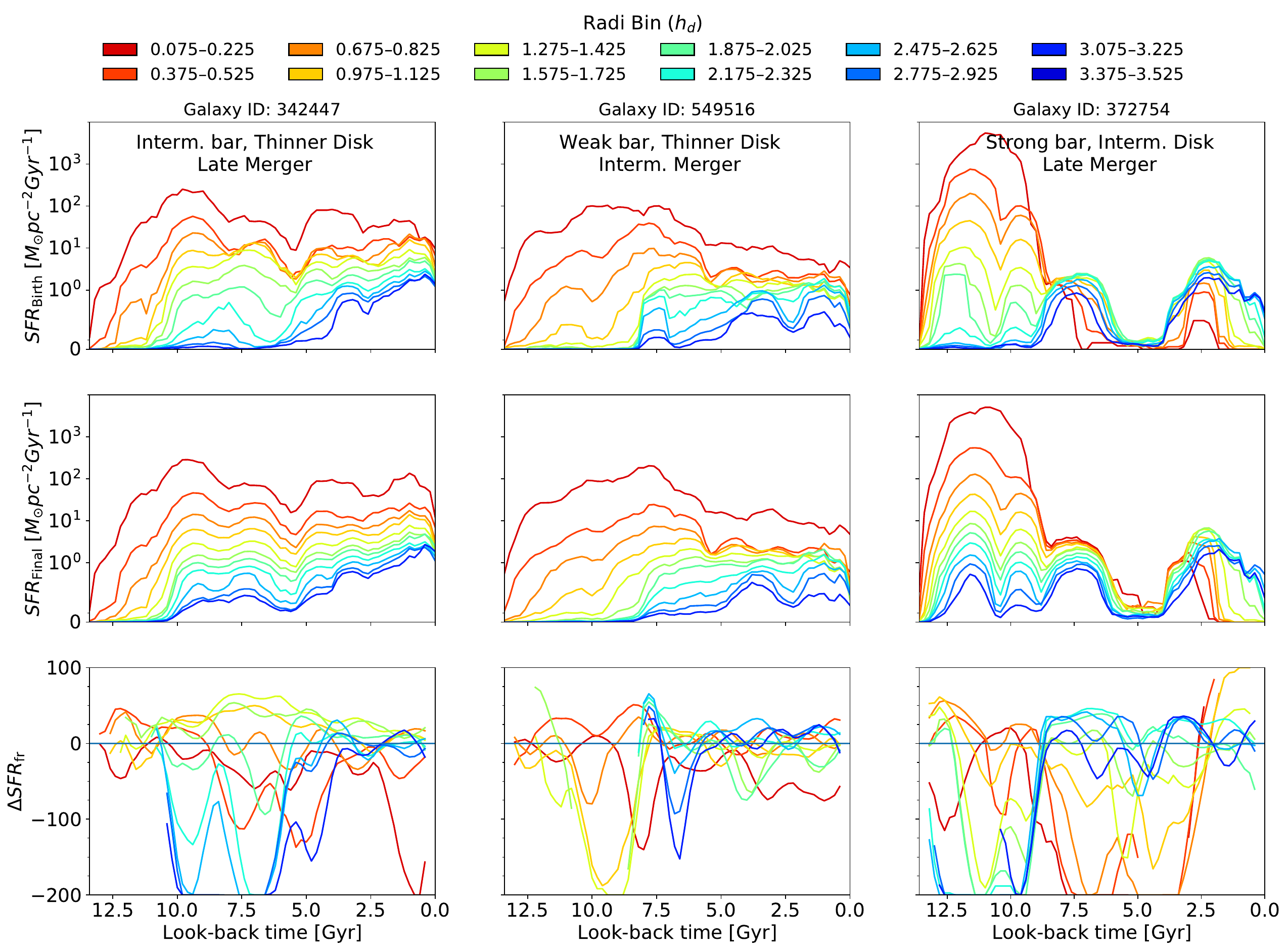}
   \caption{SFH of three different galaxies, corresponding to the three columns. The rows indicate different plots colored by the radial bins, where the top panels show the SFR$_{\text{Birth}}$ and the middle panel shows the SFR$_{\text{Final}}$. The bottom panel shows the $\Delta $SFR$_{\text{fr}}$, where the absence of a line for a given radial bin indicates that SFR$_{\text{Birth}}$ = 0 in that age bin, making the $\Delta $SFR$_{\text{fr}}$ undefined. Positive values ($\text{SFR}_{\text{Final}}$ $<$ $\text{SFR}_{\text{Birth}}$) indicate an underestimation of SFR when using final radii, while negative values show an overestimation, highlighting the impact of radial migration.}
    \label{SFH}
\end{figure*}

Figure \ref{SFH} shows the calculated values for SFR$_{\text{Birth}}$ and SFR$_{\text{Final}}$, as functions of look-back time, which measures how long ago an event occurred relative to the present day. Here, the top panels show SFR$_{\text{Birth}}$ and the middle panels show SFR$_{\text{Final}}$. By comparing these two plots, we gain insight into the impact of radial migration on the SFH of the galaxies. Since stars migrate away from their birth radii over time, binning by R$_{\text{Final}}$ instead of R$_{\text{Birth}}$ can lead to discrepancies in the estimated SFRs.
 
To quantify the relative impact of radial migration on the SFH, we computed the fractional difference, $\Delta$SFR$_{\text{fr}}$, shown in the bottom panels of Figure~\ref{SFH}, using the following equation:

\begin{equation}
\Delta \text{SFR}_{\text{fr}}=\frac{\Delta \text{SFR} \times 100}{\text{SFR}_{\text{Birth}}},
\end{equation}

where $\Delta\text{SFR}=\text{SFR}_\text{Birth}-\text{SFR}_{\text{Final}}$. To normalize variations in the SFR and avoid artificially inflated fractional differences, we imposed a threshold: the SFR in each bin was set to zero when it fell below 0.01~$\mathrm{M}_\odot,\mathrm{pc}^{-2},\mathrm{Gyr}^{-1}$. This ensures that the computed values of $\Delta \text{SFR}_{\text{fr}}$ remain meaningful and are not dominated by numerical artifacts in low-SFR regimes. A positive value of $\Delta \text{SFR}_{\text{fr}}$ indicates that using R$_{\text{Final}}$ results in an underestimation\footnote{This convention follows the definition used by \cite{minchev25} to facilitate direct comparison.} of the actual SFH, meaning that a fraction of stars have migrated away, altering the present-day interpretation of the galaxy's SFH. Conversely, a negative value suggests that using R$_{\text{Final}}$ overestimates the SFH, due to migration of stars from other parts of the disk being brought into the associated region. In cases where $\text{SFR}_{\text{Birth}} = 0$ (in $\text{M}_\odot \text{ pc}^{-2} \text{ Gyr}^{-1}$), we set $\Delta \text{SFR}_{\text{fr}}$ to NaN, as the fractional quantity becomes undefined. These instances correspond to time periods and radial regions where there was no in situ star formation, yet the final SFR is non-zero due to the presence of migrated stars. Such conditions emphasize the tangible impact of radial migration in populating parts of the galactic disk with stars that did not form locally, potentially leading to the misinterpretation of spatially resolved SFHs if migration is not properly accounted for. The color bar at the top of the figure represents different radial bins, helping to visualize how variations in SFH correlate with galactic radius.

\section{Results}
\label{sec:results}

\subsection{Individual examples}

The first column of Figure~\ref{SFH} shows the SFH for galaxy ID 342447, a thinner disk with an intermediate bar strength and a late merger. In SFR$_\mathrm{Birth}$ (top left) the SFH has peaks at $\sim9$ and $\sim5$~Gyr ago, while the outermost bin remains nearly inactive until $\sim4$~Gyr. SFR$_\mathrm{Final}$ (middle left) is much smoother: the early quiescent phase ($>6$~Gyr ago) and the 4~Gyr burst are largely washed out, especially in the outer disk. Compared to SFR$_\mathrm{Birth}$, stars originally formed in the middle bins (yellow–light green) are redistributed both inward and outward, boosting SFR$_\mathrm{Final}$ in the inner (red–orange) and outer (cyan–blue) bins. The $\Delta\mathrm{SFR}_\mathrm{fr}$ panel (bottom left) quantifies this: the intermediate inner bins are underestimated by up to $\sim50\%$, the middle disk by up to $\sim75\%$ at $\sim7$~Gyr, while the innermost bin is overestimated by up to $\sim50\%$ and the outermost by up to $\sim200\%$ once star formation begins there at $\sim11$~Gyr, confirming that the outer disk is built almost entirely from migrated stars. This confirms that the stars are redistributed from the middle disk bins to outermost regions over time. After the merger 4 Gyr ago, we see a decrease in overestimation in the middle and outer radial bins. This is because the merger-induced burst of in situ star formation in these regions begins to offset the earlier artificial overestimation caused by migrated stars, particularly in the outer disk, where star formation had been minimal or absent prior to the burst.

\begin{table*}
\caption{Galaxies showing artificial star formation in different disk regions and epochs.}
\label{tab:misleading_sfr}
\centering
\begin{tabular}{lccc}
\hline
Epoch & Outer (2.6--3.4 $\mathbf{h_d}$) & Middle (1.4--2.2 $\mathbf{h_d}$) & Inner (0.2--1.0 $\mathbf{h_d}$) \\
\hline
10+ Gyr ago    & 80\% (147) & 45\% (85) & 3\% (5) \\
8--10 Gyr ago  & 37\% (69)  & 6\% (11)  & 0\% (0) \\
6--8 Gyr ago   & 23\% (43)  & 3\% (5)   & 2\% (3) \\
$<$4 Gyr ago   & 20\% (37)  & 15\% (27) & 30\% (56) \\
\hline
\end{tabular}
\tablefoot{
A galaxy is counted in a given region and epoch if it exhibits 
SFR\textsubscript{birth} = 0 while SFR\textsubscript{final} $>$ 0, indicating that stars 
currently located in that region appear to have formed there, although they actually 
migrated from elsewhere. This illustrates how radial migration can bias spatially 
resolved star formation history estimates.
}
\end{table*}

A similar trend is seen for galaxy ID~549516 (middle column), which hosts a weaker bar and underwent a massive merger at around 7.5~Gyr ago. In SFR$_\mathrm{Birth}$ (top center) the resulting starburst is prominent in the outer disk, but it is largely absent in SFR$_\mathrm{Final}$ (middle center), indicating that these stars have migrated inward. Consequently SFR$_\mathrm{Final}$ exceeds SFR$_\mathrm{Birth}$ in the innermost bins, while the opposite holds in the outer bins. The $\Delta\mathrm{SFR}_\mathrm{fr}$ profile (bottom center) shows inner-disk overestimates of up to $\sim150\%$ and outer-disk underestimates of up to $\sim50\%$ between 13 and 7~Gyr ago, illustrating how radial migration smooths localized star-formation events over a wide radial range.

The third galaxy, ID~372754 (right column), hosts a strong bar, an intermediate disk thickness, and a late merger that produces a third SF peak beginning around 4~Gyr ago. During the first merger at $\sim7.5$~Gyr, SFR$_\mathrm{Birth}$ (top right) shows enhanced star formation predominantly in the outer disk, whereas SFR$_\mathrm{Final}$ (middle right) reveals that many of these stars have migrated inward, raising the inner-disk SFR. A similar but more radially extended pattern occurs for the second merger: SFR$_\mathrm{Final}$ suggests an almost simultaneous burst across all radii, while SFR$_\mathrm{Birth}$ shows that the inner disk lags, again highlighting the distortion caused by migration. Overall, this system reiterates the main trend that SFR$_\mathrm{Final}$ is smoother than SFR$_\mathrm{Birth}$ at look-back times $>5$~Gyr, with both the inner and outer bins gaining substantial mass from migrators.

Figure~\ref{SFH} shows that, for individual galaxies, using present-day instead of birth radii when reconstructing the SFH can bias the SFR in active regions by up to $\sim200\%$. Table~\ref{tab:misleading_sfr}  complements this by listing, for each radial range and look-back epoch, the percentage and number of galaxies that exhibit non-zero SFR$_\mathrm{Final}$ in regions where SFR$_\mathrm{Birth}=0$. This discrepancy arises because radial migration redistributes stars from their birth sites to other parts of the disk. Fractions are computed for three radial ranges: inner (0.2–1.0 $h_d$), middle (1.4–2.2 $h_d$), and outer (2.6–3.4 $h_d$), over four intervals in look-back time. To avoid misclassifying galaxies due to random fluctuations, we apply a conservative threshold: a galaxy is counted in a given region and epoch only if the condition (SFR$_\mathrm{Birth}=0$, SFR$_\mathrm{Final}>0$) is satisfied in at least three distinct age bins of width 0.2 Gyr within that epoch.

For the outer disk at early times ($>10$ Gyr ago), as seen in Table \ref{tab:misleading_sfr}, 80\% of galaxies show artificial star formation, the highest of any region or epoch. This is expected especially in the early universe as disks were still compact, and the outskirts had not yet formed stars in significant numbers. Thus, any older stars found in the outer disk at these early epochs must have migrated there from the inner regions (see Fig.~3 by \citealt{minchev25}). Over time, as disks grew inside-out and star formation propagated outward, this effect became less pronounced. By $8-10$ Gyr ago, the fraction drops to $\sim$37\%, and further to $\sim$23\% in the $6-8$ Gyr range. By the most recent epoch ($<4$ Gyr), only $\sim$20\% of galaxies show outer-disk artificial star formation, indicating that present-day outer stars are more likely to have formed in situ at late times. The 37 galaxies that exhibit no star formation in the outer disk during this epoch may reflect outer-disk quenching, where star formation has ceased entirely—possibly due to gas depletion or environmental factors that suppress star formation in the outskirts. This trend is consistent with the final-radius distributions in Figure 3 from \cite{minchev25}, which show that stars born in the inner disk (red to yellow) frequently migrate outward, populating the outer disk over time and producing the strong skewness toward larger radii seen in early epochs.

On the other hand, the middle disk in Table \ref{tab:misleading_sfr} exhibits a more transitional behavior. At $>10$ Gyr ago, nearly half ($\sim45 \%$) of galaxies show migration-driven SFH distortion. As with the outer disk, this percentage sharply declines with time, dropping to just ~6\% by $8-10$ Gyr ago and ~3\% by $6-8$ Gyr ago. Interestingly, there's a slight uptick to $\sim$15\% in the most recent epoch. This could reflect the increasing dynamical activity in the middle disk at late times due to secular processes, such as bar and spiral structure evolution. These processes may cause stars formed further out to migrate inward, though less dramatically than in the inner disk, where a larger fraction of galaxies show artificial star formation during this epoch. This supports the fact that the middle disk plays a dual role, as discussed in Section \ref{subsec:ReconstructingSFH}: at early times it receives migrating stars from the inner regions and at late times it becomes a source of outward and inward migration.

The inner disk, in the far right of Table \ref{tab:misleading_sfr}, shows the opposite trend from the outer regions. It is largely unaffected by artificial star formation at early epochs, with only $\sim$3\% affected at $>10$ Gyr and none at $8-10$ Gyr. However, its contamination steadily rises in the later epochs: $\sim$2\% at $6-8$ Gyr, and a striking $\sim$30\% of galaxies at $<4$ Gyr. This sharp increase likely reflects late-time migration of stars formed in the middle or outer disk, which are funneled inward by bar-driven dynamics, gas inflows, or disk instabilities. Such mechanisms are especially efficient at transporting stars to the inner galaxy during the last few Gyr, leading to the buildup of central mass and a mismatch between true and inferred SFH.

\subsection{Influence of bar strength on $ \Delta \text{SFR}_{\text{fr}}$} \label{subsec:SFHBarStrength}

To better understand the effect of radial migration for different galaxies, we analyze $\Delta$SFR$_{\text{fr}}$ across six bar strength bins. We apply the same radial and vertical cuts as in Section \ref{SFH_comparison}, but with larger radial and age bins to smooth the plots and reduce noise. Specifically, we define six radial bins with a bin size of 0.30$h_d$, centered at every 0.60$h_d$ maintaining a maximum radial cut at 4.5$h_d$. For age binning, we use 2.5 Gyr bin widths, measured at every 0.5 Gyr. The galaxies are then categorized based on their bar strength at $\mathit{z=0}$, with the smallest bin covering $0.0 < A_2/A_0 < 0.1$ and the largest bin including galaxies with $A_2/A_0 > 0.5$, with increments of $A_2/A_0 = 0.1$, for six bar strength bins.

\begin{figure*}
   \centering
   \includegraphics[width=2.0\columnwidth]{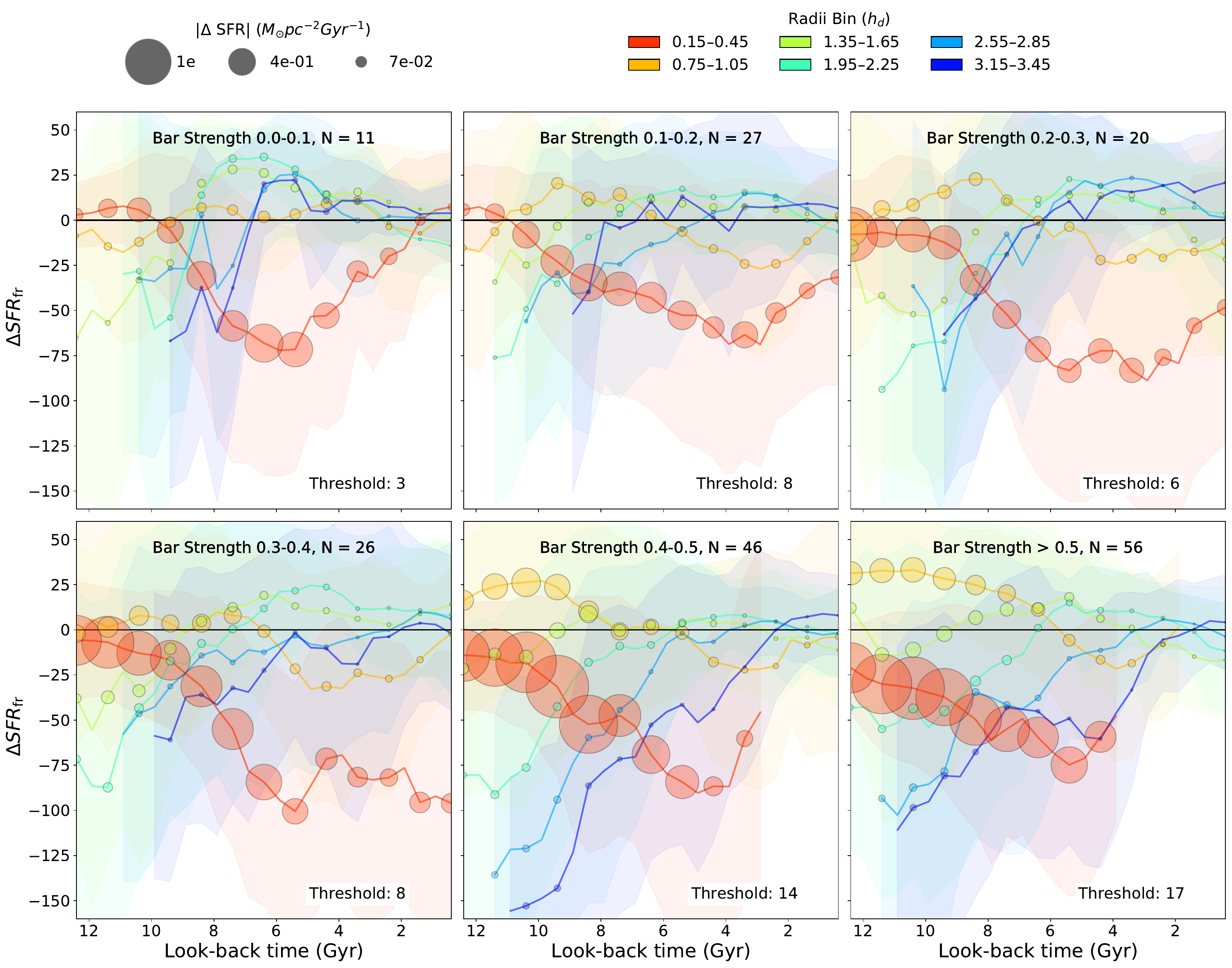}
   \caption{Average $\Delta$SFR$_{\text{fr}}$ for galaxies grouped by bar strength at $\mathit{z=0}$. Each panel shows a different bar strength bin (increasing from top to bottom), with colored lines indicating radial bins in units of disk scale-length ($h_d$). Shaded circles denote the magnitude of $\Delta$SFR within the corresponding age bin, illustrating the amount of displaced mass across the disk, while the shaded area reflects the standard deviation. A completeness threshold is also applied, where a point is shown only if at least 35\% of galaxies in that bar strength group have nonzero SFR\textsubscript{Birth} in the corresponding age bin.}
    \label{SFH_BarStrengths}
\end{figure*}

Figure \ref{SFH_BarStrengths} presents the average $\Delta$SFR$_{\text{fr}}$ for galaxies categorized by bar strength, with the size of each data point corresponding to $|\Delta\text{SFR}|$, which is the total displaced mass, within the given age bin. This visualization allows us to analyze how the bar strength influences the estimation of the SFR across different radial regions and epochs. To ensure the reliability of the trends shown for each bar strength bin, we impose a completeness threshold: the average is only computed for a given age bin if more than 35\% of the galaxies in that group have nonzero SFR\textsubscript{Birth} values in that bin, ensuring that only star-forming regions contribute to the computed averages. This criterion helps avoid bias from under-sampled bins and ensures that the displayed trends reflect representative behavior across the galaxy subsample.

Focusing first on the innermost radial bin (represented in red) in Figure \ref{SFH_BarStrengths}, we observe a clear trend at early times (before 9 Gyr ago), where increasing bar strength generally leads to a shift from underestimation to overestimation of the SFR by up to 40\%. For galaxies with the stronger bars ($0.5 < A_2/A_0 < 0.6$), such as the MW \citep{Khoperskov2024}, this trend even increases the overestimation of the SFR by up to 75\% after 9 Gyr ago. This overestimation is due to stars that were originally born in the intermediate inner region (orange line) that later migrated inward along bar-driven orbital resonances. As a result, the present-day concentration of stars in the center gives the impression that significant star formation occurred there early on, even when it did not. This highlights how strong bars can amplify the impact of radial migration, complicating efforts to reconstruct the true SFH based solely on stellar distributions. The large point sizes in the innermost regions, especially in high bar strength bins, further emphasize this effect by indicating a greater amount of displaced stellar mass.  Notably, we also find that in galaxies with stronger bars, the innermost radial bin shows a halt in star formation activity as early as 4 Gyr ago, suggesting that bar driven gas dynamics may suppress late time central star formation. This is consistent with observations of barred galaxies, where bars are found to quench central star formation by driving gas inward and stabilizing the inner disk \citep{Geron2024, Renu2025}. It is also in line with TNG50 results showing that strong, slowly rotating bars are preferentially hosted by centrally quenched disks affected by kinetic-mode AGN feedback \citep{semczuk24}.

In contrast, the intermediate inner disk (orange line) exhibits the opposite behavior at early times. Here, increasing bar strength correlates with a reduction in SFR overestimation, eventually leading to an underestimation of up to 30\% in the strongest barred galaxies. This divergence from the innermost radial bin's (red line) trend can be attributed to the kinematics of stars within strongly barred systems. Specifically, in these galaxies, stars in the innermost disk likely become trapped in the x2 orbital family—closed elliptical orbits that exist within the inner Lindblad resonance (ILR) \citep{Athanassoula1992,Pfenniger1984}. These orbits prevent stars that are in the innermost radial bin from migrating outward, while the bar-spiral interaction facilitates inward migration to this bin, particularly along the leading edge of the bar \citep{Marques2024}. As a result, stars just beyond the innermost radial bin we considered are funneled inward, while those already within this region remain confined, unable to migrate outward. This dynamic redistribution of stellar populations explains the contrasting SFR estimation trends observed between the innermost and intermediate inner radial bins in strongly barred galaxies. At later times (within the past 4 Gyr), the behavior shifts. The intermediate inner radial bin begins to show an overestimation of SFR by up to 30\%, while the innermost radial bin in galaxies with weaker bars—where star formation still persists—exhibits the opposite trend, showing a reduction in overestimation.

Moving to what we considered the middle disk (green to cyan lines) in Figure \ref{SFH_BarStrengths}, which typically corresponds to the region where spiral arms are most prominent, we observe a distinct trend in $\Delta$SFR$_{\text{fr}}$. At early times (before $\sim$9 Gyr ago), the middle disk shows a growing overestimation of the SFR by up to 50\% with increasing bar strength. This implies that stars currently found in this region were likely born further inward and migrated outward over time, a process enhanced by stronger bars. Notably, the onset of significant star formation in the middle disk lags behind that of the inner disk, indicating an inside-out growth pattern common in disk galaxies. In the more recent epochs (past $\sim$9 Gyr in look-back time), this early overestimation transitions to underestimation of up to 30\%, which is lesser for galaxies with stronger bars. This shift suggests that while early migration may have populated the middle disk with stars from inner regions, later star formation occurred more in situ. The reduction in underestimation over time indicates that star formation in the middle disk has increasingly aligned with the present-day stellar distribution, making the migration effects less pronounced in recent epochs.

\begin{figure*}
   \centering
   \includegraphics[width=2.0\columnwidth]{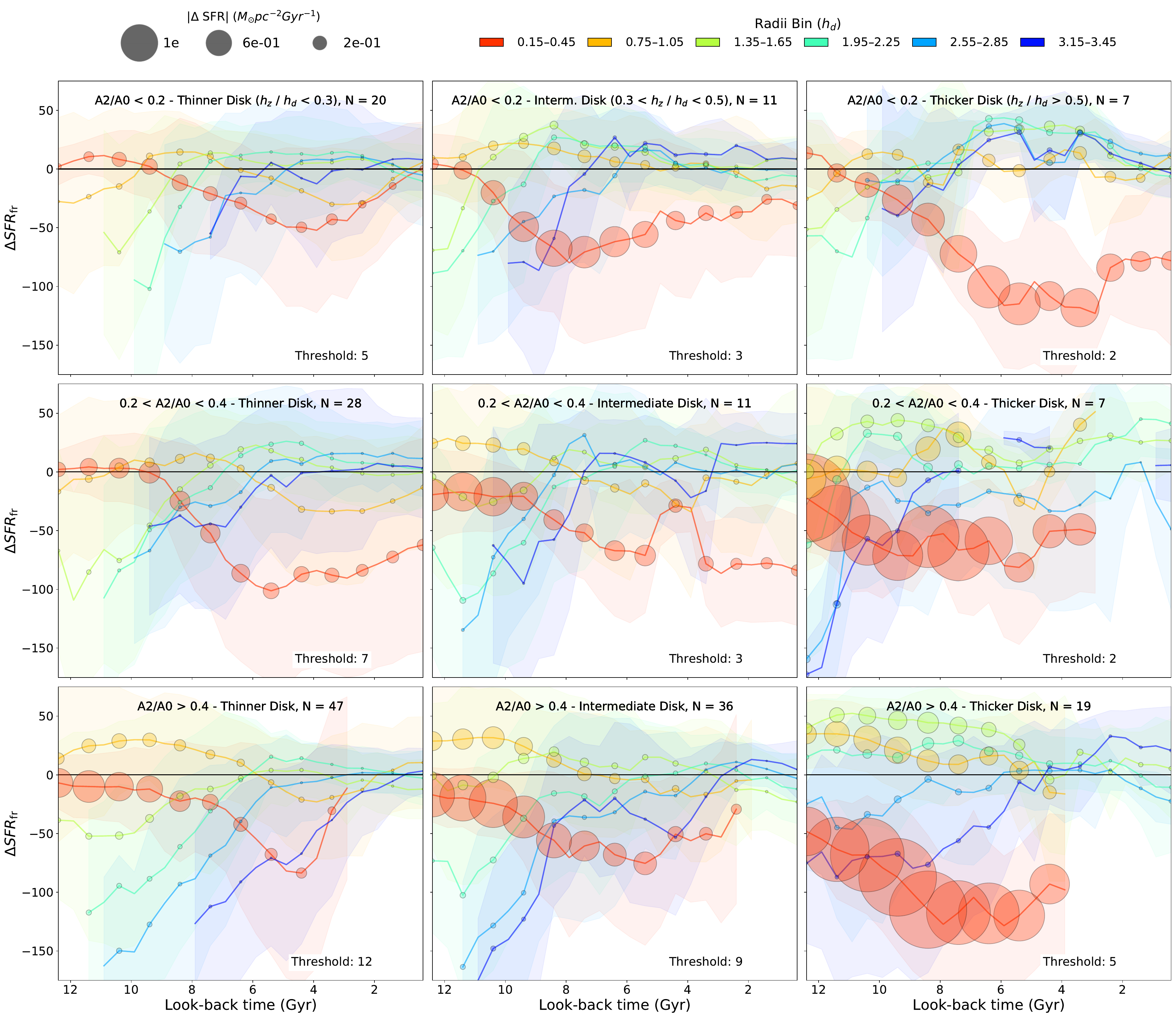}
   \caption{Mean $\Delta$SFR$_{\text{fr}}$ for galaxies grouped by bar strength (rows) and disk thickness (columns), shown as a function of look-back time across radial bins. Colored lines represent SFHs in six radial bins ($h_d$), and shaded circles mark the magnitude of $\Delta$SFR, reflecting migration-induced differences between final and birth radii. From left to right, columns correspond to increasing disk thickness, while from top to bottom corresponds to increasing bar strength. A completeness threshold is applied, where a point is shown only if at least 35\% of galaxies in that disk thickness group have nonzero SFR\textsubscript{Birth} in the corresponding age bin.}
    \label{SFH_DiskThickness}
\end{figure*}

A similar pattern is observed in the outermost radial bin (blue lines), where star formation tends to begin later than the innermost radial bin, consistent with the inside-out growth of disk galaxies \citep{Matteucci1989,Matteucci2021}. We also see that star formation in the outermost bin begins earlier for stronger bars. Starting around 10 Gyr ago, a pronounced overestimation of SFR, up to 150\%, emerges in the outer disk, especially in galaxies with stronger bars. For these strongly barred systems, the overestimation persists throughout cosmic time, suggesting a continuous influx of stars into the outer regions — likely driven by radial migration from the middle disk. In contrast, weakly barred galaxies exhibit a more complex behavior: while they also show early overestimation, a temporary underestimation of up to 25\% is visible around 7 Gyr ago, implying that some stars formed in the outer disk may have migrated inward during this period, reducing the mismatch between birth and final radii. 

The sustained overestimation in the outermost disk of strongly barred galaxies indicates that stellar populations initially formed in the middle disk were redistributed outward over time. This outward migration is consistent with the results from \cite{Baba2025}, who found that during bar formation, stars formed in the inner disk migrate outward, especially into the outer disk beyond the outer Lindblad resonance (OLR), where local star formation is minimal. This follows the idea that strong bars can drive angular momentum exchange and facilitate the outward transfer of stars through resonant interactions or bar-spiral coupling \citep{Minchev2010,Minchev2011,hilmi20, Marques2024}. As bars grow stronger and more dynamically influential, their ability to shape stellar orbits extends farther out well beyond their length due to the effect of their Outer Lindblad Resonance (OLR) interacting with spiral resonances, reshaping the spatial distribution of stars and amplifying the discrepancies in SFR reconstructions based solely on present-day locations.

In general, increasing bar strength correlates with a greater degree of overestimation in the outer disk, suggesting that the migration effects induced by bars become increasingly dominant toward larger radii. A summary of the effects of radial migration with increasing bar strength, by radial bin, is provided in Table \ref{EffectsTable}. In addition to the average trends, the shaded regions around each radial bin curve represent the standard deviation of $\Delta \text{SFR}_{\text{fr}}$ across galaxies in each bar strength bin. These ribbons provide insight into the diversity of migration histories among galaxies with similar bar strengths. At early times (look-back time $>$ 8 Gyr), the spread is wider—particularly in the outer and inner disk bins—indicating significant variation in the extent and timing of migration. As time progresses toward the present, these shaded regions narrow, especially in the outer disk, implying that the SFHs in these regions become more homogeneous across the sample since stars do not move as much at this time. This trend reflects a convergence in migration efficiency and a stabilization of disk dynamics in the more recent evolutionary phases.

\subsection{Influence of disk thickness on $ \Delta \text{SFR}_{\text{fr}}$}
\label{subsec:SFHDiskThickness}

Now that we have examined the effects of radial migration for different bar strengths, we extend our analysis to investigate its impact in conjunction with disk thickness. To ensure consistency, we use the same radial bin sizes, bin centers, and age bins as in our previous analysis. Additionally, we apply a completeness threshold of 25\%, which requires that at least one-quarter of the galaxies in each disk thickness subgroup contribute valid star formation values within a given age bin to be included in the averaged results. This threshold ensures statistical reliability while accounting for the smaller number of galaxies in each disk thickness bin, as compared to the bar strength bins. Here, we categorize galaxies into three distinct bar strength groups: weakly barred galaxies, where $A_2/A_0 < 0.2$; moderately barred galaxies, with $0.2 < A_2/A_0 < 0.4$; and strongly barred galaxies, where $A_2/A_0 > 0.4$. Similarly, we classify galaxies based on disk thickness using the ratio $h_{z,\text{thick}}/h_d$, with thinner disks defined as $\le 0.30$, intermediate as $0.30$–$0.46$, and thicker disks as $\ge 0.46$, as introduced in Section \ref{subsubsec:DiskThickness}. From these classifications, we analyze how both bar strength and disk thickness influence radial migration and SFH.

Figure \ref{SFH_DiskThickness} presents $\Delta$SFR$_{\text{fr}}$ as a function of bar strength and disk thickness. The figure consists of nine panels, where from top to bottom, bar strength increases, while from left to right, disk thickness increases. The plot clearly demonstrates the strong influence of bar strength on radial migration, as discussed in Section \ref{subsec:SFHBarStrength}. Interestingly, the innermost radial bin (red line) in strongly barred thicker galaxies (bottom right) tends to cease forming stars earlier than strongly barred thinner galaxies (bottom left), at around 2–4 Gyr prior. This is seen in Figure \ref{SFH_DiskThickness}, where the mean $\Delta$SFR$_\text{fr}$ is undefined, which means that no stars are formed within the threshold—suggesting a stronger quenching of central star formation. This suppression may result from bar-driven gas inflows that deplete the central gas reservoir, or from enhanced vertical heating, which stabilizes the inner disk and inhibits further gas collapse. Within this innermost region, a clear trend emerges at look-back times greater than 6 Gyr: galaxies with thicker disks (right) show a stronger overestimation of the SFR by up to 125\% compared to thinner disks (left), particularly in systems with stronger bars. While this could be mistaken as evidence that thicker disks experience more efficient radial migration, further analysis (see Appendix \ref{app:SFH_BirthComp}) reveals that this trend is largely driven by higher intrinsic SFR$_\text{Birth}$ in the innermost regions of these galaxies. In other words, the thicker-disk galaxies simply formed more stars in their inner disks (both red and orange lines) early on, resulting in a larger population available to migrate, which then exaggerates the apparent overestimation that we see in the innermost radial bin when using final stellar positions. The visibly larger shaded points in this region reflect this higher amount of displaced stellar mass, but do not imply a more efficient migration mechanism in thicker disks. Instead, the redistribution of these larger stellar populations, especially in the innermost radial bin, reflects the action of early bar-driven migration processes. 

Thick disks often host older, dynamically hotter stars that formed more centrally (including those now in the intermediate inner bin, orange), allowing some outward migration over time. In strongly barred systems this is further enhanced, as bar–spiral torques efficiently redistribute stellar angular momentum in the inner disk, amplifying both inward and outward migration \citep{Marques2024}. For this bin, however, the overall migration pattern is set mainly by bar strength, with only a weak dependence on disk thickness.

\begin{figure*}
   \centering
   \includegraphics[width=2.0\columnwidth]{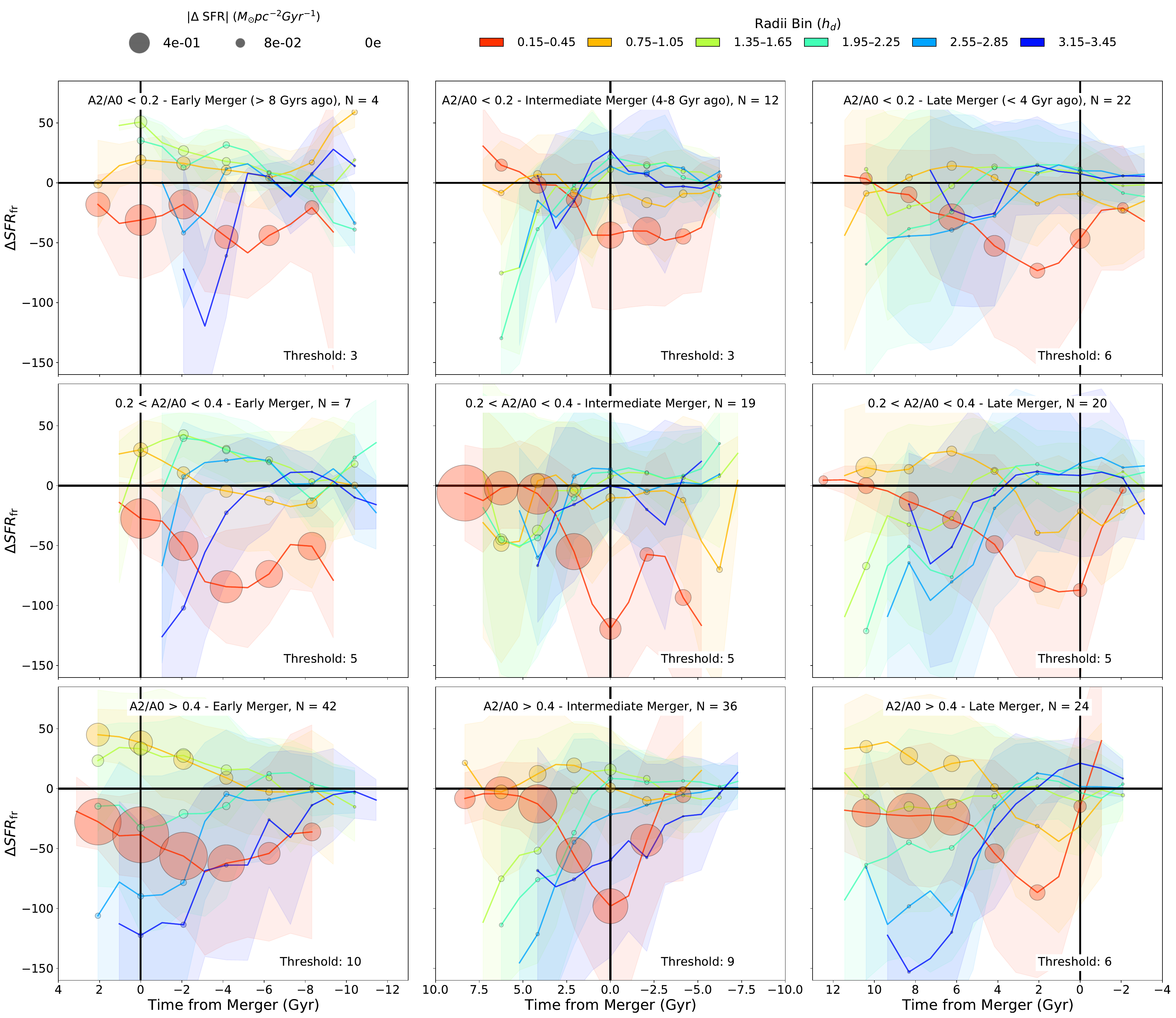}
   \caption{Mean $\Delta$SFR$_{\text{fr}}$ for galaxies grouped by bar strength (rows) and merger history (columns), plotted against look-back time for different radial bins. Each column shows a distinct merger epoch: early ($>$ 8 Gyr ago), middle ($4-8$ Gyr ago), and late ($<$ 4 Gyr ago). The x-axis is centered on the time of the merger, with 0 marking the merger event. Negative values correspond to look-back time before the merger, and positive values indicate time after the merger, measured in Gyr. Colored lines correspond to six radial bins in units of $h_d$, and shaded circles represent the magnitude of $\Delta$SFR, highlighting the extent of radial migration before and after mergers. A completeness threshold is applied, where a point is shown only if at least 35\% of galaxies in that merger history group have nonzero SFR\textsubscript{Birth} in the corresponding age bin.} 
    \label{SFH_DiskMerger}
\end{figure*}

From the middle (green–cyan) to outer (blue) bins, thinner disks (left) show systematically larger $|\Delta\mathrm{SFR}_\mathrm{fr}|$ than thicker disks (right). Thus, when using final positions, the outer-disk SFR is overestimated by up to $\sim160\%$ in strongly barred galaxies and $\sim100\%$ in weakly barred ones. This reflects the lower vertical velocity dispersions of thinner disks \citep{Elmegreen2006}, whose kinematically cold stars are more susceptible to bar- and spiral-driven resonant scattering, so that many stars migrate from the middle disk into the outskirts. In contrast, dynamically hotter thick disks, with more eccentric and randomly oriented orbits, respond less efficiently and therefore migrate less. In these systems, especially with strong bars, $\Delta\mathrm{SFR}_\mathrm{fr}$ in the outer disk is smaller at look-back times $>6$~Gyr, because early star formation already populated these regions, reducing the apparent migration-induced bias.

The behavior of the middle disk also depends on thickness. In thick disks, at early times ($\gtrsim 6$ Gyr) the intermediate inner and middle bins together host much of the stellar mass formed in the disk, and stars are redistributed from this combined region both inward and outward. In thin disks, by contrast, a larger fraction of stars are born in the intermediate inner bin and then migrate both inward into the innermost bin and outward into the middle and outermost bins. As a result, the middle disk of thin galaxies shares many of the migrator-dominated properties of the outer disk, whereas in thick-disk galaxies the middle and outer bins remain more distinct. These trends are summarized in Table~\ref{EffectsTable}.

The shaded ribbons reflect the standard deviation in $\Delta$SFR$_\text{fr}$, capturing the scatter among galaxies in each bar strength and disk thickness group. A key trend observed across many panels is that the spread tends to be larger at early times (look-back time $>$ 8 Gyr), particularly for the inner and outer disk regions. This indicates that the diversity in migration and early SFHs is greater in the early universe, likely due to varying merger histories, gas accretion rates, and the timing of bar formation. Thicker disk galaxies generally show wider ribbons, especially in the intermediate, and outer disk regions in the early times, suggesting a broader range of migration histories and star formation evolution. In contrast, thinner disk galaxies tend to display narrower ribbons at later times, particularly in the outer disk, indicating more consistent and less varied migration effects among these systems. This narrowing of the spread with decreasing look-back time also supports the idea that galaxies evolve toward more stable and uniform SFH profiles in the later stages of their development.

\subsection{Influence of merger history on $\Delta \text{SFR}_{\text{fr}}$}
\label{subsec:SFHMerger}

To investigate how merger events shape radial migration, we categorize galaxies by the timing of their most massive merger, identified from the SubLink merger trees (see Section \ref{subsubsec:Mergers}). Figure \ref{SFH_DiskMerger} presents the average $\Delta \text{SFR}_\text{fr}$ across merger epoch and bar strength. Each column corresponds to the epoch of the most massive merger: early (mergers $> 8$ Gyr ago), middle ($4-8$ Gyr ago), and late ($< 4$ Gyr ago), while rows indicate increasing bar strength. As before, colors denote different radial bins, and shaded points represent the magnitude of $\Delta$SFR in each bin and age slice. A threshold of 25\% completeness is applied per age bin, as in Section \ref{subsec:SFHDiskThickness}, to ensure reliable averages. This is adjusted to 35\% for bins that have less than 10 galaxies, due to the existence of bins with very few galaxies. The plot shows that the trends with merger epoch are generally subtle and less pronounced than those associated with bar strength and disk thickness. A closer inspection reveals distinct differences in the radial migration direction and origin of migration that correlate with merger timing. 

In early massive mergers (left panels), the disk is still in the process of assembling, and the merger-driven burst of star formation occurs predominantly in the innermost radial bin (red line) and intermediate inner (orange line) to middle (green lines) radial bins. At these early times, the outer disk remains largely unpopulated, and stellar migration originates mainly from the intermediate inner to middle disk—evident from the plot where these radial bins are underestimated—indicating that more stars were born in these regions than are present there at the final time. In contrast, the innermost (red line) and outer (blue lines) bins are overestimated for all bar strength bins up to 10 Gyr after the merger, implying that these regions received stars via migration. While the innermost region hosts much of the early star formation, it contributes relatively little to outward migration—especially in galaxies with strong bars, where bar-driven dynamics tend to confine stars to the center. Instead, stars formed slightly farther out, specifically in the intermediate inner and middle radial bins, tend to migrate both inward and outward, establishing the foundation of the disk's radial structure. The inward migration causes a consistent overestimation in the innermost radial bin by up to 75\%, while the outward migration shows an overestimation in outer disk by up to 120\%. This bidirectional redistribution suggests that early mergers play a critical role in shaping the initial configuration of the galaxy. In addition to this, for the outer radial bins, we find that increasing bar strength enables earlier outer disk formation, with stars beginning to populate the outer regions as early as 2 Gyr before the merger, compared to weakly barred galaxies where outer disk formation is delayed and only becomes significant up to 2 Gyr after the merger. 

\begin{table*}
\renewcommand{\arraystretch}{1.2}
\centering
\caption{Effects of bar strength, disk thickness, and merger history on SFH distortions.}
\label{EffectsTable}
\begin{tabularx}{\textwidth}{%
  >{\raggedright\arraybackslash}p{0.13\textwidth}%
  >{\raggedright\arraybackslash}X%
  >{\raggedright\arraybackslash}X%
  >{\raggedright\arraybackslash}X}
\hline
Disk region & Increasing bar strength & Increasing disk thickness & Merger history effects \\
\hline
Innermost disk (red)\par\vspace{1.5ex}0.15--0.45 $h_d$ &
Age $>9$ Gyr: SFR overestimated by up to $\sim75\%$; strong bars funnel old stars inward and can quench central SF at late times. &
Age $>6$ Gyr: SFR overestimates stronger in thicker disks, up to $\sim125\%$; reflects more centrally concentrated old pops. rather than more efficient migration. &
Early massive mergers ($>8$ Gyr ago) drive inward migration from larger radii, producing inner-disk SFR overestimates up to $\sim75\%$. \\
\hline
Intermediate inner disk (orange)\par\vspace{1.5ex} 0.75--1.05 $h_d$ &
Age $>6$ Gyr: SFR underestimated by up to $\sim30\%$; Age $<6$ Gyr: SFR overestimated by up to $\sim30\%$ as bars feed this region from both sides. &
Only weak dependence on thickness; the age trends are largely set by bar strength, with little variation as disks get thicker. &
Early massive mergers ($>8$ Gyr ago) lead to SFR underestimates because stars formed here migrate both inward (to the center) and outward (to the middle disk). \\
\hline
Middle disk (green)\par\vspace{1.5ex} 1.35--2.25 $h_d$ &
Age $>9$ Gyr: SFR overestimated by up to $\sim50\%$; Ages $\sim2$--9 Gyr: SFR underestimated by up to $\sim30\%$ as region transitions from being populated by inward-born migrators to more in situ stars. &
For old populations (Age $>6$ Gyr), SFR can be underestimated by up to $\sim50\%$ in thicker disks relative to thin disks, reflecting less efficient migration into the middle disk. &
Intermediate-time mergers (4--8 Gyr ago) sustain middle-disk SFR underestimates through two-way redistribution of stars both inward and outward. \\
\hline
Outer Disk (blue)\par\vspace{1.5ex} 2.55--3.45 $h_d$ &
Age $>6$ Gyr: SFR overestimated by up to $\sim150\%$; strong bars continuously feed the outer disk with stars migrated outward from inner radii. &
Maximum SFR overestimate occurs in thin disks: up to $\sim160\%$ for strongly barred and $\sim100\%$ for weakly barred galaxies; the bias decreases as disks become thicker. &
Early mergers ($>8$ Gyr) cause an overestimation of SFR  by up to $\sim120\%$; late mergers ($<4$ Gyr) underestimate SFR by up to $\sim20\%$ due to enhanced inward migration from outermost disk. \\
\hline
\end{tabularx}
\tablefoot{
“Age” refers to stellar age at $z=0$ (equivalently, lookback time since formation).
The table summarizes how bar strength, disk thickness, and merger history modify
the inferred star formation histories (SFHs) in different radial regions by altering
the balance between in situ star formation and radial migration-driven redistribution.
}
\renewcommand{\arraystretch}{1.0}
\end{table*}

Intermediate massive mergers (middle column) display a transitional behavior between early and late mergers. In strongly barred galaxies (bottom center), the mergers trigger a star formation burst primarily in the middle disk (green to cyan lines), whereas in weakly barred galaxies (top center), the star formation peak occurs in the outer regions (blue lines). This is determined based on the observed overestimation in the corresponding radial bins at the time of the merger. Specifically, the middle disk shows a significant overestimation in strongly barred galaxies, while the outer disk bins show overestimation in weakly barred galaxies, indicating that the majority of stars currently residing there formed during the merger epoch and migrated afterward. The contrast between the intermediate massive mergers for strongly barred and weakly barred galaxies likely stems from the differing ability of bars to regulate gas dynamics: strong bars redistribute gas from the outer to the inner and middle disk over time, often depleting gas from outer disk star before the merger occurs \citep{Khoperskov2018}. As a result, strongly barred galaxies have lower intrinsic star formation in the outer disk than the amount of stars migrating towards it, concentrating merger-induced star formation in the middle disk. In contrast, weakly barred systems retain more gas in their outer regions, allowing merger-triggered star formation to occur in the outskirts, with many of these stars subsequently migrating inward. The extent and direction of this migration depend on where the star formation occurs. In strongly barred galaxies, stars formed in the middle disk tend to migrate both inward and outward, leading to an overestimation of SFR in the outer disk by up to 60\% at the time of the most massive merger. In weakly barred galaxies, where star formation is concentrated farther out, the resulting inward migration causes an underestimation of outer-disk SFR by up to 25\%. This spatial dependence also affects the innermost radial bin: when stars are formed in the middle disk, such as for strongly barred galaxies, the resulting overestimation in the innermost region reaches up to 120\%, whereas if star formation occurs in the outer disk as in weakly barred galaxies, the overestimation is more moderate—around 45\%.

In late massive mergers (right panels), a distinct pattern emerges: both before and at the time of the most massive merger, the innermost radial bin (red line) shows an overestimation of SFR by up to 90\%, while the outer radial bins (blue lines) exhibit an underestimation of up to 20\%. This suggests that the merger-induced star formation is concentrated in the outer disk. However, because our outermost radial bin spans 3.15–3.45 $h_d$, we are primarily sensitive to stars migrating inward. The observed trends imply that many stars formed in the outer disk during the merger epoch are later displaced inward due to dynamical perturbations associated with the merger. This inward shift is less pronounced in galaxies that formed their disks earlier (often those with stronger bars), likely because such systems are more dynamically stable by the time of the merger and therefore more resistant to significant redistribution. Unlike the bidirectional migration observed in intermediate mergers—where the middle disk acts as a mixing zone—late mergers appear to cause more unidirectional migration, predominantly inward, from the outer to the inner disk. It should also be noted that because late massive mergers occur only within the last 4 Gyr, the contribution of bar-driven migration is already significant prior to the merger. Strongly barred galaxies (bottom right) promote greater stellar redistribution even before the merger event, as evidenced by the large overestimation seen in the outer radial bins, including the outermost bin with overestimation reaching up to 150\%, and in the innermost bin up to 90\%. This indicates that prior to the merger, bar-induced migration already reshaped the radial distribution of stars, and the subsequent merger further amplifies this inward migration trend. A summary of the key trends in SFH distortions across radial bins as a function of merger history is presented in Table \ref{EffectsTable}.

In addition, the shaded ribbons in the figure highlight the variability in radial migration effects among galaxies with different merger histories. In early mergers (left panels), the standard deviation is relatively narrow at early epochs, suggesting that galaxies in this group follow a more uniform trend in migration and SFR redistribution. This may reflect a more coherent formation history, where stars formed early and migrated in predictable ways. In contrast, intermediate mergers (center panels) exhibit the widest ribbons, especially in the middle and outer radial bins (green to blue lines), pointing to significant diversity in how galaxies respond to mergers during this transitional epoch. This large spread implies a range of possible merger masses relative to disk formation, leading to more heterogeneous migration signatures. Late mergers (right panels) show modest spread at early times for inner disk but growing divergence near the merger epoch (2–4 Gyr look-back), especially in galaxies with weaker bars. The spread is lower in strongly barred galaxies, possibly due to earlier disk settling and stronger internal dynamical control.

\section{Conclusions}
\label{sec:Conclusion}

We investigated how radial migration distorts the spatially resolved SFHs of MW and M31 analogs using 186 well-aligned disk galaxies from the TNG50 cosmological simulation. By comparing SFHs derived from stellar birth radii with those derived from present-day positions, we quantified both the magnitude of migration-induced biases and their spatiotemporal structure, and we examined how these biases vary with bar strength, disk thickness, and merger history.

A striking and recurrent pattern is the appearance of artificial star formation in regions that did not actually form stars at those times. Most notably, 80\% of galaxies seem to have outer-disk stars older than 10~Gyr, well before those regions began forming stars, indicating that these populations must have migrated from radii interior to the outer disk. Similar effects occur in 45\% of galaxies at intermediate radii during early epochs, and in 30\% of galaxies in quenched inner disks over the past 4~Gyr, revealing a widespread redistribution of stellar populations across time. Migration also smooths the SFH, washing out localized bursts and suppressions by spreading stars into neighboring radial bins.

The strength and radial signature of these distortions depend strongly on galaxy structure and evolutionary history. Strong bars drive large SFR overestimates in both the innermost (up to 75\%) and outermost (up to 150\%) disk through resonance-driven redistribution, while thinner disks---being dynamically colder---are more susceptible to outer-disk overestimates of up to 160\%. Thicker disks show large inner-disk overestimates (up to 125\%) due to early central star formation creating large populations available to migrate. Merger timing strongly shapes the migration patterns: early mergers promote bidirectional migration, intermediate mergers show bar-dependent redistribution, and late mergers feed outer-disk star formation that later migrates inward. 

Our results underscore that failing to account for stellar migration can lead to severe misinterpretations of when and where stars formed. This has direct consequences for studies of chemical evolution and for the reconstruction of SFHs in both the MW and external galaxies. 

For the MW, the biases quantified here are especially relevant because precise chemical, kinematic, and age information is already available from Gaia \citep{gaia18,drimmel23} and large spectroscopic surveys (e.g., APOGEE; \citealt{majewski17}), and will be further improved by ongoing and upcoming programs such as SDSS-V, WEAVE, and 4MOST \citep{sdss25,dalton12,dejong19,walcher19}, together with asteroseismic missions (K2, TESS, PLATO; \citealt{howell14,ricker15,rauer14}). In this context, several studies have proposed using stellar chemistry and ages to infer birth radii in the Milky Way \citep{minchev18, feltzing19, Frankel2020, lu24, ratcliffe23, dantas25,ratcliffe25}, and thereby correct local SFH estimates for the effects of radial migration.

This work, together with \citet{minchev25}, establishes the scale of migration-induced SFH biases in cosmological simulations across bar strength, disk thickness, and merger history. On the observational side, \citet{Ratcliffe2025b} apply the orbit-superposition method of \citet{khoperskov25} in combination with the stellar birth-radius estimation technique of \citet{Ratcliffe2025} to APOGEE data, making explicit use of the chemo-kinematic constraints provided by Gaia and large spectroscopic surveys. This methodology allows the orbit-mass-weighted SFH of the MW to be traced back to the true stellar formation sites while consistently accounting for stellar mass loss. Together, these studies provide a cohesive framework for interpreting SFHs, linking cosmological simulations, detailed migration modeling, and the observational reconstruction of our Galaxy’s formation history.

\begin{acknowledgements}
We thank the anonymous referee for a constructive report that improved the clarity and interpretation of this work. We also thank F. Hammer and M. Semczuk for helpful comments. IM and BR acknowledge support by the Deutsche Forschungsgemeinschaft under the grant MI 2009/2-1. This work is based on the Master’s thesis of J.P. Bernaldez (2025, Leibniz Institute for Astrophysics Potsdam, AIP).
\end{acknowledgements}

\bibliography{biblit2}{}
\bibliographystyle{aa}

\begin{appendix}
\section{Disk centering and alignment}\label{app:Centering}

The TNG50 simulation uses a comoving Cartesian coordinate system, in which the positions of stellar and gas particles are given relative to a coordinate grid that expands with the universe. The origin of the simulation is located at the corner of the simulation box. To analyze individual galaxies, it is necessary to re-center each galaxy by transforming the initial coordinate system into a galactocentric coordinate system, for each MW and M31 subhalo.  Thus, the comoving coordinates were converted by multiplying the comoving distance to a scale factor $a$, which depends on the redshift, given by $a=1/(1+z)$. Then, an approximate center of each galaxy was determined using the minimum gravitational potential of all the stellar particles in the galaxy for each snapshot. To more precisely align the galactic center, all stellar particles within 1 kpc of the minimum gravitational potential were identified, and the mean of their coordinates was calculated. The coordinates of all stellar particles were then rotated so the galaxy appears face-on in the $x-y$ plane and edge-on in the $x-z$ plane, aligning its orientation with the angular momentum vector. This reorientation removes distortions from arbitrary viewing angles and ensures that radial distances are physically meaningful, enabling more accurate analyses of structure, dynamics, and migration history. For each snapshot, we selected stars within a radius $R_{max}$ specific to that galaxy. $R_{max}$ was determined by visually inspecting the stellar density profile and selecting a value that encompasses the main disk structure while excluding loosely bound or satellite material. We then computed the eigenvectors from their covariance matrix, and applied a rotation matrix based on the sorted eigenvectors. The galactic center was determined independently for each galaxy and epoch, and the corresponding rotation matrix was used to align the stellar distribution.

\section{Calculating disk scale-length and scale-height}
\label{app:ScaleLengths}

\begin{figure}
   \includegraphics[width= \columnwidth]{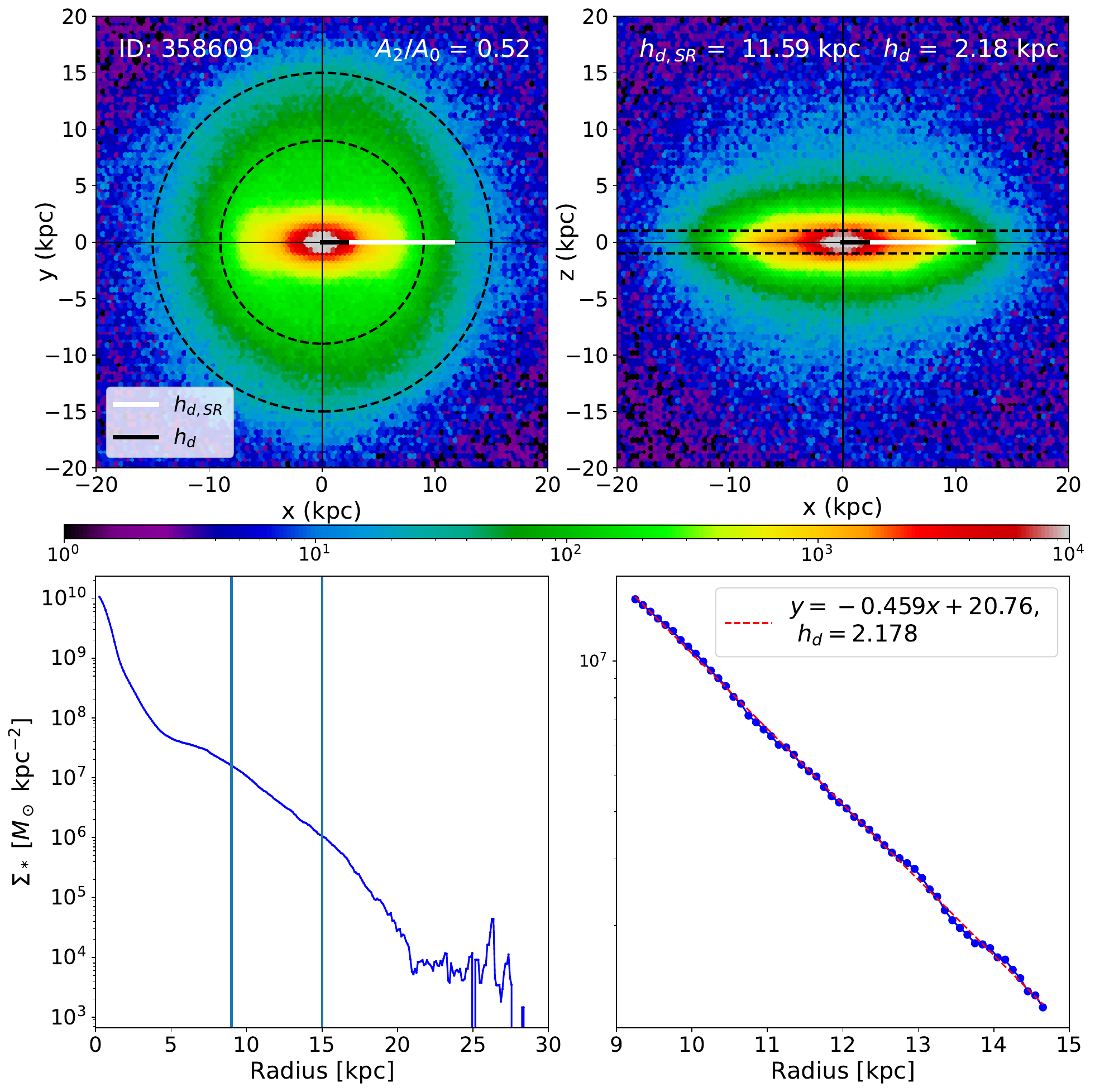} 
   \caption{Example of linear profile fitting for disk scale-length calculation. The top panels show the galaxy face-on (left) and edge-on (right), with white and black lines indicating $h_{d,SR}$ and the $h_d$ we measured, respectively. The bottom panels display the surface density profile ($\Sigma_*$): with vertical lines marking the fitting radius, while the right panel shows the profile within this radius. Our recalculated $h_d$ provides a more accurate measurement of the disk scale-length by better excluding bulge-dominated regions and outer disk noise.}
   \label{Fig_DiskScaleLength} 
\end{figure}

Scaling the positional coordinates of the galaxies relative to their disk scale-length is essential for comparing trends when averaging over a large sample of galaxies with varying sizes. The disk scale-lengths of the galaxies in the TNG50 simulation were calculated in \cite{SotilloRamos2022}, $h_{d,SR}$, by fitting an exponential profile to the radial stellar surface density distribution in face-on projection within one to four times the half-mass radius. However, for some galaxies, the $h_{d,SR}$ was very large compared to the actual size of the galaxy. For example, one of the galaxies with an unusually large $h_{d,SR}$ is shown in Figure \ref{Fig_DiskScaleLength}, clearly demonstrating that the $h_{d,SR}$ (in white, top panels) spans the entirety of the galactic disk, at 11.59 kpc, which should not be the case. Moreover, the $h_{d,SR}$ values are significantly larger than expected for the MW, with $h_d$ of about 2–3 kpc \citep{Bovy2013,Sackett1997}, and M31, with estimates around 5.3 kpc \citep{Courteau2011}. To address these limitations, we recalculated the disk scale-lengths by examining each galaxy individually with the aim to produce a good fit to the stellar density in the disk region, that is excluding the bulge and halo. This approach allows for a more accurate representation of the disk's true structure, especially in galaxies where the $h_{d,SR}$ is overestimated, and helps tailor the analysis to the specific characteristics of each galaxy.

\begin{figure}
   \includegraphics[width= 0.7\columnwidth]{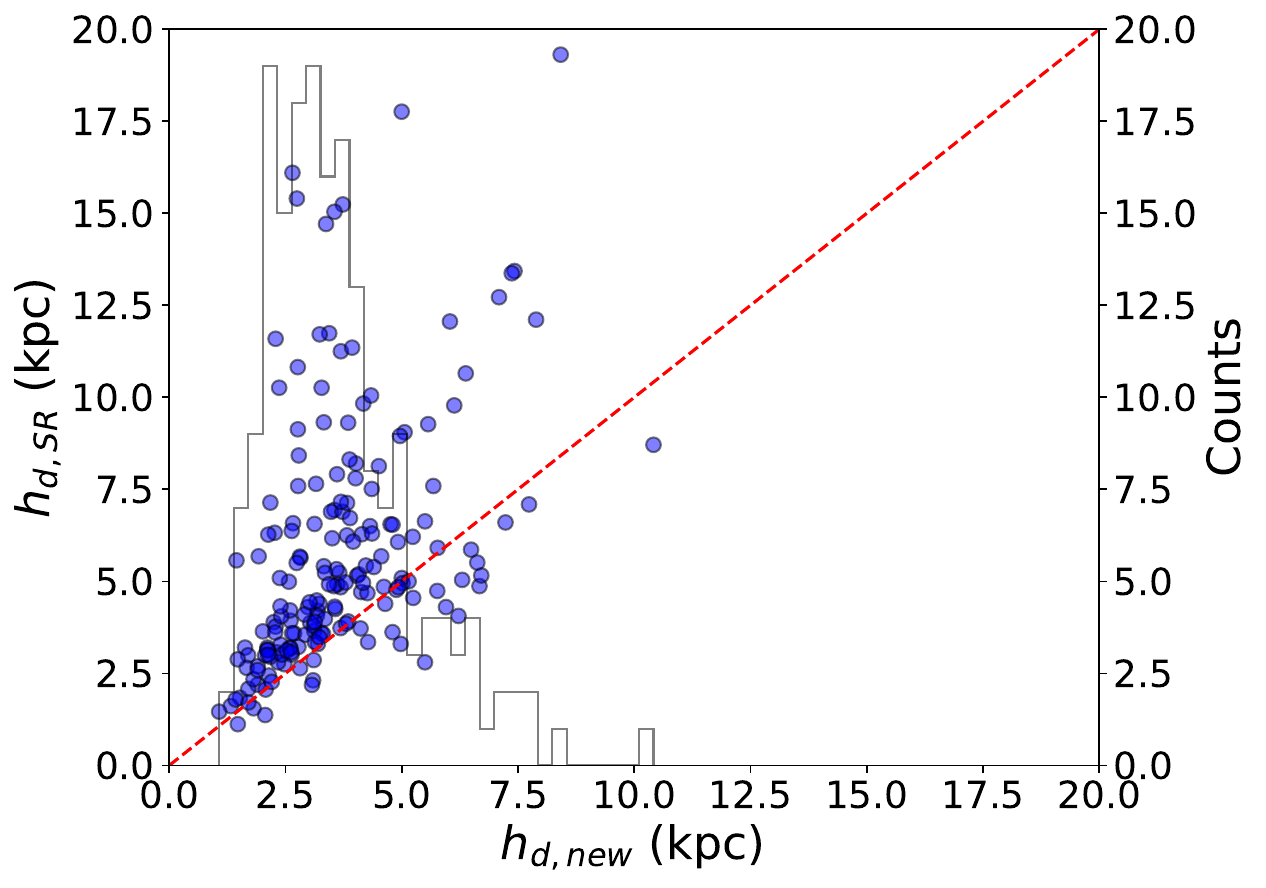} 
   \caption{Comparison between the disk scale-lengths $h_{d,SR}$ from \cite{SotilloRamos2022} and our recalculated values $h_d$ using individual exponential fits. The red dashed line indicates the 1:1 relation. While both methods show broad agreement at lower values, $h_{d,SR}$ tends to systematically overestimate the disk scale-length compared to our refined measurements. The overestimation becomes more pronounced at larger values. A histogram of $h_d$ is shown in gray to indicate the distribution of recalculated values.}
   \label{DiskScaleLength_Comparison} 
\end{figure}

\begin{figure*}
   \centering
   \includegraphics[width= 1.5\columnwidth]{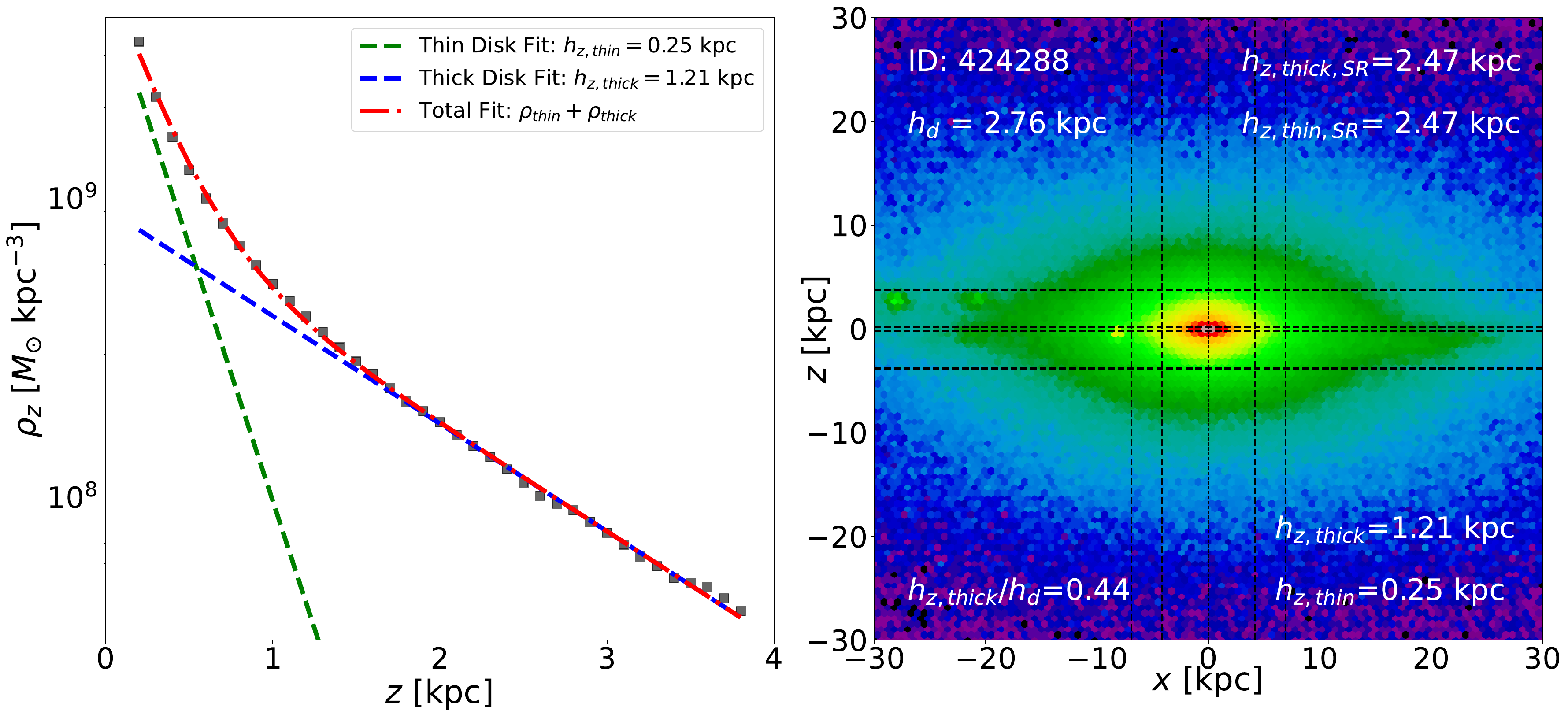}
   \caption{Example of the double exponential profile fitting of the vertical density ($\rho_z$) of a galaxy. The left panel shows the thin disk fit (green line), thick disk fit (blue line), total fit (red line), and data points represented by square markers. The right panel displays the edge-on view of the galaxy, with black lines indicating the regions where the fitting was performed.}
   \label{DisKScaleHeight} 
\end{figure*}

\begin{figure}
   \centering
   \includegraphics[width= 0.8\columnwidth]{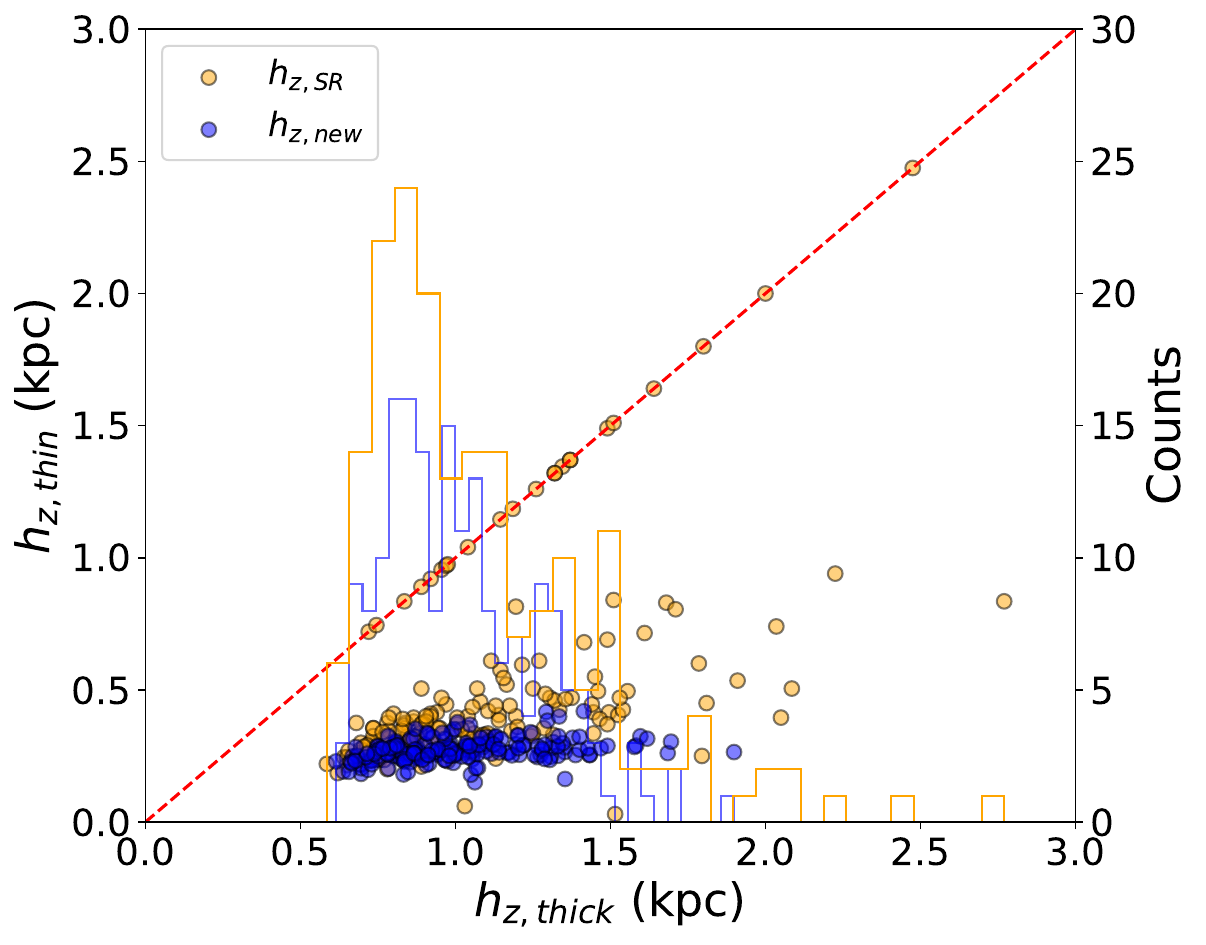}
   \caption{Comparison of $h_{z,thick}$ and $h_{z,thin}$ using our fitted values (blue) and those from \cite{SotilloRamos2022} (orange), with a 1:1 reference line in red. Our measurements yield consistently smaller $h_{z,thin}$, especially at larger $h_{z,thick}$, suggesting better separation of thin and thick disk components. }
   \label{DiskScaleHeight_Comparison} 
\end{figure}

The disk scale-length calculations were carried out by fitting a linear profile to the log-scale of the radial stellar surface density distribution, excluding the bulge region and, when a Type II or III break was present, the outer disk beyond the break radius, at redshift 0. To accurately analyze the structural properties of the galaxy and produce a more representative radial profile, specific cuts were applied to the stellar population. Stars that were not born within the galaxy were excluded. A vertical cut was also applied by including only stars within $|z_{\text{Birth}}|<1 kpc$ and $|z_\text{Final}|<1$ kpc, where $z_{\text{Birth}}$ and $z_\text{Final}$ refer to the star's vertical position at birth and at the final simulation time respectively to exclude those associated with the halo or other extended structures. The radial surface density profile was constructed by binning the stellar mass in concentric annuli with a width of 0.5 kpc, measured from the center of the galaxy. From this, we fitted the linear profile only beyond the bulge and bar region, which is characterized by a steeper slope at the galaxy's inner radius. This region was identified individually for each galaxy by assessing the surface density profile. 

An example disk scale-length calculation for a galaxy can be found in Figure \ref{Fig_DiskScaleLength}. Here, the top left plot shows the face-on view of the galaxy, with the dashed black circular annuli indicating the radial region where the linear fit was made. Similarly, the top right plot shows an edge-on view of the galaxy, with the dashed black horizontal lines indicating the cuts made in $z$, and displaying both $h_{d,SR}$ (in white), and our measured disk scale-length, $h_{d}$ (in black). The bottom left panel shows the surface density of stars with ages less than 13 Gyr, while the bottom right panel shows a zoomed surface density plot within the fitting radius. 
  
This process was repeated for all the galaxies used in the study, with the radial regions used varying depending on the galaxy's surface density profile. Figure \ref{DiskScaleLength_Comparison} shows a comparison of $h_{d,SR}$ and $h_{d}$. The red dashed line shows the 1:1 relation, indicating where the two measurements would be equal. From the scatter plot, it is immediately clear that the $h_{d,SR}$ values are systematically larger for a wide range of galaxies. Many points lie above the 1:1 line, particularly at intermediate and high disk scale-lengths, suggesting that the method used in \cite{SotilloRamos2022} tends to overestimate disk size. 
  
This discrepancy likely stems from the fact that $h_{d,SR}$ was computed over a wide radial range (1–4 times the half-mass radius), which can encompass regions like the central bulge, outer disk, or even satellite structures. In contrast, most of the $h_{d,new}$ values cluster between 2 and 5 kpc, consistent with observational estimates \citep{Bovy2013, Courteau2011}, which shows that the use of a more localized fit for each galaxy that excludes the inner bulge and focuses specifically on the disk part produces more consistent scale-length estimates that are better suited for comparing galaxy structure and radial migration for different systems.
   
The disk scale-height was also calculated to identify and compare thinner and thicker disk galaxies. Here, ``thinner'' or ``thicker'' refers to the ratio between the disk scale-height and scale-length. In contrast, the terms ``thin'' or ``thick'' disks refers to distinct structural components within spiral galaxies. A thicker disk galaxy is one in which the vertical scale-height is relatively large compared to its radial scale-length, while thinner disk galaxies have relatively smaller vertical scale-heights. Stars near the midplane ($|z_\text{Final}| < 0.2$ kpc) are more numerous due to higher stellar mass density, but the vertical profile in a narrow radial bin flattens in this region, making it poorly described by a simple two-component exponential model. To improve accuracy, we excluded $|z_\text{Final}| < 0.2$ kpc from the fits, following the approach of \cite{Minchev2015}. The vertical density of the galaxies was calculated by binning star particles in the vertical direction with 0.2 kpc-wide, non-overlapping bins from $|z_\text{Final}| = 0.2$ to 4 kpc to exclude the central region, and then smoothed using a 0.1 kpc rolling average to reduce noise in the profile. Additionally, the vertical density was only calculated within the radius of 1.5-2.5 $h_d$, ensuring that the analysis is focused on the middle disk, to avoid contamination from a central bulge, a boxy peanut structure, or disk flaring in the outer disk and halo. The vertical stellar density was computed with the double-exponential profile of the form:
\begin{equation}
   \rho(z) = \rho_{thin}\times \exp\left(\frac{-|z_\text{Final}|}{h_{z, thin}}\right)+\rho_{thick}\times \exp\left(\frac{-|z_\text{Final}|}{h_{z, thick}}\right),
\end{equation}
which captures the contributions from both the thin and thick disk components. The first term represents the thin disk, where the density decreases with a scale-height $h_{z,thin}$, while the second term corresponds to the thick disk, which has a larger scale-height $h_{z,thick}$. 

\section{SFH$_{\text{Birth}}$ comparison}\label{app:SFH_BirthComp}

\begin{figure*}[ht]
   \centering
   \includegraphics[width=1.5\columnwidth]{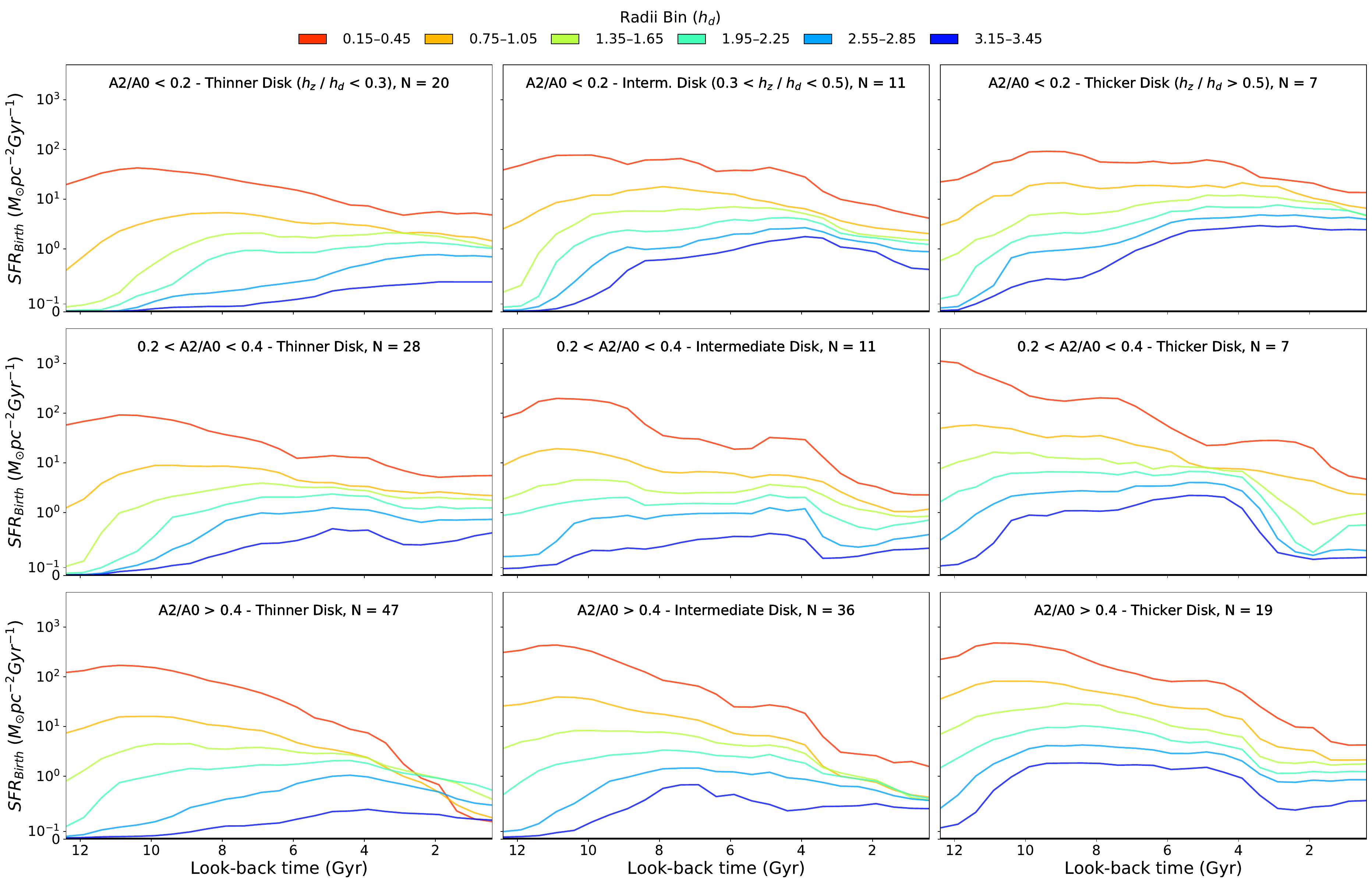}
   \caption{Average SFR$_\text{Birth}$ as a function of look-back time for galaxies grouped by bar strength (rows) and disk thickness (columns). Each colored line represents the six radial bins in terms of $h_d$. Galaxies with stronger bars tend to exhibit earlier and more centrally concentrated star formation, while those with thicker disks generally show higher SFR$_\text{Birth}$ for the same radial bin.}
    \label{SFH_DiskThickness_SFRB}
\end{figure*}

Figure \ref{SFH_DiskThickness_SFRB} presents the average SFR$_\text{Birth}$ across look-back time for galaxies classified by both bar strength and disk thickness. Each panel represents a specific bin in bar strength (rows) and disk thickness (columns), allowing a systematic comparison across galaxy structural types. The SFRs are color-coded by radial distance from the galactic center (in $h_d$), from red (innermost) to blue (outermost). Across all panels, star formation predominantly occurs in the inner disk at early times (look-back time $>8$ Gyr). When comparing by disk thickness, as in Section \ref{SFH_DiskThickness}, galaxies with thicker disks generally exhibit higher SFR$_\text{Birth}$ in the innermost and intermediate bins during early epochs, which allows for more mass to be displaced by migration. Moreover, thicker disk galaxies also show more pronounced early star formation in the outer disk, unlike their thinner counterparts, which exhibit delayed outer-disk activity. This trend supports the conclusion that outer disks in thicker galaxies formed earlier than in thinner disks, potentially due to dynamical heating or earlier gas accretion histories.

\end{appendix}

\end{document}